# Symmetric Capacity of the Gaussian Interference Channel with an Out-of-Band Relay to within 1.15 Bits

Ye Tian, *Student Member, IEEE,* and Aylin Yener, *Member, IEEE*

*Abstract*—This work studies the Gaussian interference channel (IC) with a relay, which transmits and receives in a band that is orthogonal to the IC. The channel associated with the relay is thus an out-of-band relay channel (OBRC). The focus is on a symmetric channel model, in order to assess the fundamental impact of the OBRC on the signal interaction of the IC, in the simplest possible setting. First, the linear deterministic model is investigated and the sum capacity of this channel is established for all possible channel parameters. In particular, it is observed that the impact of OBRC, as its links get stronger, is similar to that of output feedback for the IC. The insights obtained from the deterministic model are then used to design achievable schemes for the Gaussian model. The interference links are classified as extremely strong, very strong, strong, moderate, weak, and very weak. For strong and moderate interference, separate encoding is near optimal. For very strong and extremely strong interference, the interference links provide side information to the destinations, which can help the transmission through the OBRC. For weak or very weak interference, an extension of the Han-Kobayashi scheme for the IC is utilized, where the messages are split into common and private. To achieve higher rates, it is beneficial to further split the common message into two parts, and the OBRC plays an important role in decoding the common message. It is shown that our strategy achieves the symmetric capacity to within 1.14625 bits per channel use with duplexing factor 0.5, and 1.27125 bits per channel use for arbitrary duplexing factors, for all channel parameters. An important observation from the constant gap result is that strong interference can be beneficial with the presence of an OBR.

*Index Terms*—Interference Relay Channel, Out-of-Band Relay, Approximate Capacity, Deterministic Model, Nested Lattice Codes

## I. INTRODUCTION

Broadcast and superposition are two features unique to the wireless medium. Interference is an inevitable consequence of these two features, and is a crucial factor that impacts the capacity of wireless networks. Interference channel (IC), which consists of two source-destination pairs, is the simplest

Manuscript received October 29, 2010; revised July 16, 2011; accepted March 05, 2012. This work was supported by the National Science Foundation under Grant CNS 0716325, and DARPA ITMANET Program under Grant W911NF-07-1-0028. This work was presented in part at the Forty-eighth Annual Allerton Conference On Communication, Control, and Computing [1], and ICC International Conference on Communications, ICC'11 [2].

The authors are with the Department of Electrical Engineering at the Pennsylvania State University, University Park, PA 16802 (email: yetian@psu.edu, yener@ee.psu.edu).

model that characterizes the effect of interference in a network, and is thus a basic building block for wireless ad hoc networks. References [3]–[9] established the capacity for the IC when the interference is either strong or weak. However, for the general case, the capacity is open.

Relay channel (RC) is another important building block for wireless networks. It has been shown that the relay can cooperate with the source to increase the transmission rate of a point to point channel [10]–[12]. The capacity of RC is also established for special cases and the general case remains open [10].

Recent efforts [13]–[18] introduce a relay node in the IC setting, resulting in a new fundamental channel termed the interference relay channel (IFRC). In the IFRC, the relay can perform *signal relaying* [16], [18] as in the traditional relay channel, *compute-and-forward* [18], [19] or *interference forwarding* [13], [14]. All the schemes can help increase the achievable (sum) rate of the IC under different channel conditions.

Recent references [15], [18] derived sum rate upperbounds, which complement each other, for the Gaussian interference relay channel (GIFRC). The capacity region of IFRC is only known for special cases [13], [15]. For the general IFRC, the capacity region is open, since it inherits the challenges of both IC and RC, with increased signal interaction. To simplify the channel model and understand the fundamental effect of signal relaying and interference forwarding, reference [17] proposed a model where the relay operates in bands orthogonal to the underlying IC, termed therein the *interference channel with an out-of-band relay* (IC-OBR). For IC-OBR, reference [17] first considered the case when the links associated with the relay are all orthogonal to each other, and obtained capacity results for some channel configurations. A more general model, where only the incoming links and outgoing links of the relay are orthogonal, is also considered in [17]. The channel model for this case is shown in Figure 1, which contains an underlying IC, and the sources and the destinations have access to another band orthogonal to the IC. The communication between sources and destinations in the band orthogonal to the IC is only possible with the help of a relay, which is termed the out-of-band relay (OBR). The relay is half-duplex, i.e., the incoming links of the relay are orthogonal to its outgoing links, either in time or in frequency. We call the channel associated with the relay the out-of-band relay channel (OBRC). The sources and the destinations have access to both the IC and the OBRC. Reference [17] established the optimality conditions of signal relaying and interference forwarding with separable or non-separable encoding between IC and OBRC. The resulting strategies do achieve the sum capacity for certain channel parameters. On the other hand, they can also be far from outerbounds for some other channel parameters. It is desirable to gain a fundamental understanding to the impact of an out-of-band relay on the signal interaction and the capacity in this model for all channel settings. This is the main goal of our work. For simplicity, we use IC-OBR to refer to the model shown in Figure 1, since this is the channel model investigated in this work.

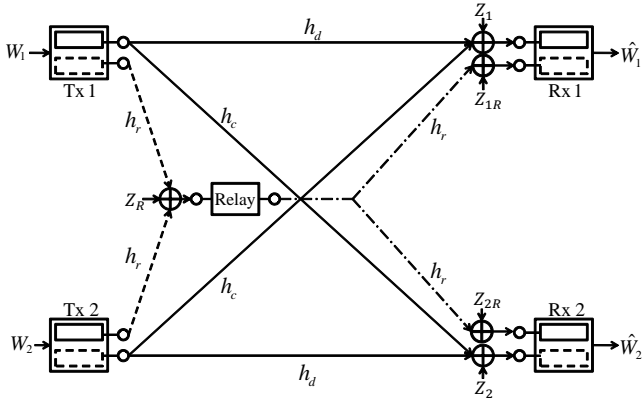

Fig. 1. The Gaussian interference channel with an out-of-band half-duplex relay.

To make our presentation self-contained, we provide a detailed introduction of the motivation and characteristics of IC-OBR, although a similar discussion can be found in reference [17]. In practice, this model can describe an OFDM based wireless network, where some subcarriers experience large path loss or frequency selective fading and need to be assisted by a relay, or a wireless local area network (WLAN) with some short range radio, such as Bluetooth, enabled for relaying data. This model simplifies the signal interaction, but is still general enough for us to assess the impact of cooperation and interference on capacity: It *physically* separates the relayed signals and the interfered signals, but keeps the possible *statistical* correlation between them.

We focus on the symmetric channel, where the channel gain of two direct links, two interference links and links associated with the relay are assumed to be equal, respectively. This simplified setting retains the essence of what we set out to accomplish, i.e., the impact of the relaying scheme and its interaction with interference, without having to accommodate the difference between channel gains when studying the capacity. We first study the linear deterministic model using the approach developed in [20]. The deterministic model allows us to focus on the interaction of the signals by eliminating the noise at the receiver. This approach is also utilized in [21]–[24] to obtain approximate capacity results for various channel models.

For the symmetric deterministic IC-OBR, we characterize the sum capacity for all possible channel configurations. We observe that the presence of the OBR impacts the capacity in a manner similar to that observed in the presence of output feedback for the IC, see [21]. The essence lies in that the available resources, i.e., signal spaces, can be better utilized using the OBRC. For the converse, we derive outerbounds via the aid of judiciously designed genie information. For achievability, we first observe that for the sum capacity optimal transmission strategies for the deterministic IC, some signal spaces are left unused to avoid interference. Using the out-of-band relay (OBR), we show that these signal spaces can be utilized. For the case when the interference link is stronger than the direct link, we further classify the interference as *strong*, *very strong* and *extremely strong*. When the interference is strong, it is optimal for the sources to transmit independent information bits through the IC and the OBRC, that is, separate encoding is optimal. When the interference is very strong or extremely strong, the interference links can carry additional information bits, which serve as side information to help the decoding of the signal transmitted from the OBRC. For the cases when the interference link is weaker than the direct link, we further classify the interference as *moderate*, *weak* and *very weak*. When the interference is moderate, separate encoding between the IC and the OBRC is optimal. When the interference is weak or very weak, we use the unused signal spaces of the IC to transmit new information bits, which causes interference at the destinations. The OBRC can now be utilized to remove the interference. Overall, for all possible cases, we show that the achievable sum rates match the outerbounds. We further show that, in fact, the full capacity region can be characterized when the interference is strong.

We next utilize the insights obtained from the deterministic model to construct achievable strategies for the Gaussian channel. For the achievable strategy, we use a combination of nested lattice codes [25] and Gaussian codes for the OBRC, and Gaussian codes for the IC. For strong interference, separate encoding is optimal, similar to the deterministic model. When the interference is very strong or extremely strong, the sources can transmit some additional messages through the interference links. We align the signals carrying these messages at noise level at the direct links. With the OBRC, we show that these messages can be recovered by the intended destinations to achieve within a constant gap of the outerbounds. In particular, when interference is extremely strong, the channel acts as if there are two disjoint OBRC helping each source-destination pair.

When the interference is moderate, separate encoding between the IC and the OBRC with Han-Kobayashi (HK) strategy at the IC results in achievable rates that are within a constant gap of the outerbounds. When the interference is weak or very weak, the sources also use HK strategy for the IC, where the messages are splitted into common and private, and the private messages are aligned at noise level at the interference links. The common message is the primary source of interference at the non-intended receiver. From the perspective of the receiver, we call the common message from the interferer the *common interference message*, and the common message from the intended source the *common information message*. Without the OBR, both common information and interference messages must be decoded from the IC at all time, and this approach achieves within 1 bit of the sum capacity for the IC [7]. This approach, however, does not work well for the IC-OBR. With the OBR, we show that it is beneficial to further split the common messages into two parts for weak interference. Both parts of the common information message are decoded from the IC, while the common interference message is decoded jointly from the IC and OBRC. For very weak interference, however, the sources do not need to further split the common messages. The common information messages are still decoded from the IC, but the common interference messages are decoded from the OBRC. By deriving new outerbounds, we show that our scheme achieves rates that are *within 1.14625 bits of the symmetric capacity with duplexing factor 0.5*, and *1.27125 bits* of the symmetric capacity with arbitrary duplexing factors. An important observation from the constant gap result is that strong interference can be *beneficial* in improving the capacity with the presence of an OBR. This observation shows *the positive effect of strong interference*, whereas for IC without OBR, strong interference at most has a *neutral effect*, i.e., it does not reduce capacity.

The remainder of the paper is organized as follows: Section II describes the channel models. Section III derives the outerbounds for the linear deterministic model based on a genie-aided approach, describes the achievable schemes and presents the sum capacity results for the linear deterministic model. Section IV presents outerbounds and achievable strategies for the Gaussian channel, and the constant gap result. Section V concludes the paper.

## II. SYSTEM MODEL

### A. The symmetric Gaussian interference channel with an out-of-band relay (Gaussian IC-OBR)

The Gaussian IC-OBR is shown in Figure 1, which consists of a two-user interference channel (IC), i.e., two pairs of sources and destinations, and a relay operating in orthogonal bands, i.e., an out-of-band relay (OBR). The OBR is half-duplex and thus uses part of its frequency band to receive signals and the remainder to transmit signals. The sources and destinations operate in a common band which forms the interference channel. The relay for cooperation helps the transmitters via its incoming band and the receivers via its outgoing band. We consider the symmetric case, where for the interference channel, the gain of the direct link is $h_d$ and the gain of the interfering link is $h_c$. The gain of the links associated with the relay is $h_r$.

To communicate to its the destination, source $i$ encodes a message $W_i \in \{1, 2, \cdots, 2^{mR_i}\}$ into a set of codewords $\{X_i^m, X_{iR}^{\alpha m}\}$, where $X_i^m$ is the codeword to be sent into the IC while $X_{iR}^{\alpha m}$ is the codeword to be sent into the OBRC, and $\alpha$ is the duplexing factor. Note that if $\alpha m$ is not an integer, the effect of rounding to its nearest integer on the achievable rate is negligible, as $m \to \infty$. We assume separate power

constraints on the IC and OBRC:

$$\frac{1}{m} \sum_{t=1}^{m} E[X_{i,t}^2] \le P_i, \tag{1}$$

$$\frac{1}{\alpha m} \sum_{t=1}^{\alpha m} E[X_{iR,t}^2] \le P_{iR}. \tag{2}$$

The relay generates codewords based on the signals received from its incoming bands, i.e., $X_{R,\alpha m+1}^m = f_R(Y_R^{\alpha m})$ with power constraints

$$\frac{1}{(1-\alpha)m} \sum_{t=\alpha m+1}^{m} E[X_{R,t}] \le P_R. \tag{3}$$

The channel outputs for the IC are

$$Y_{1,t} = h_d X_{1,t} + h_c X_{2,t} + Z_{1,t} \tag{4}$$

$$Y_{2,t} = h_c X_{1,t} + h_d X_{2,t} + Z_{2,t} \tag{5}$$

for $t = 1, \cdots, m$. The channel outputs at the relay are

$$Y_{R,t} = h_r X_{1R,t} + h_r X_{2R,t} + Z_{R,t} \tag{6}$$

for $t = 1, \cdots, \alpha m$. The channel outputs for the OBRC are

$$Y_{1R,t} = h_r X_{R,t} + Z_{1R,t} \tag{7}$$

$$Y_{2R,t} = h_r X_{R,t} + Z_{2R,t} \tag{8}$$

for $t = \alpha m+1, \cdots, m$. With out loss of generality, we assume $P_i = P_{iR} = P_R = 1$, and $Z_{j,t}$ $(j = 1, 2, 1R, 2R, R)$ are independent, unit variance Gaussian random variables.

The symmetric capacity is defined as

$$C_{sym} = \sup\{R : (R, R) \in \mathcal{C}\}, \tag{9}$$

where $\mathcal{C}$ is the capacity region. For the symmetric channel, the rate points that maximize the sum rate achieve the symmetric capacity. We thus focus on the sum capacity of this channel.

As a first step, we investigate the deterministic model to find the optimal transmission strategy, which provides us with insights about the signal interactions in the Gaussian channel. The deterministic model is described in the next section.

### B. The deterministic symmetric interference channel with an out-of-band relay (IC-OBR)

The deterministic IC-OBR is shown in Figure 2, where for the interference channel, the gain of the direct link is $n_d$ and the gain of the interfering link is $n_c$. The gain of the links associated with the relay is $n_r$. $n_d, n_c, n_r$ are integers.

Let $w_1 \in \{1, 2, \ldots, 2^{mR_1}\}$, $w_2 \in \{1, 2, \ldots, 2^{mR_2}\}$ denote the messages of the two sources. Each transmitter uses an encoding function $x_i : w_i \to \mathbb{F}_2^q \times \mathbb{F}_2^q$ $(i = 1, 2)$ with $q = \max\{n_d, n_c, n_r\}$, to generate codewords $\mathbf{x}_i^m(w_i) = [\mathbf{x}_{i1}^m, \mathbf{x}_{i2}^{\alpha m}]$, where $\alpha$ is the duplexing factor, and

$$\mathbf{x}_{ik} = [x_{ik,1}, x_{ik,2}, \ldots, x_{ik,q}]^T, k = 1, 2 \tag{10}$$

$$x_{ik,m} \in \mathbb{F}_2, m = 1, 2, \ldots, q. \tag{11}$$

The OBR sends $\mathbf{x}_{r,\alpha m+1}^m$ to the destinations using the outgoing bands. The signal $\mathbf{x}_{r,\alpha m+1}^m$ is generated based on the signals received from the incoming bands of the OBR in the past, i.e., $\mathbf{x}_{r,\alpha m+1}^m = f(\mathbf{y}_r^{\alpha m})$, where $\mathbf{x}_r \in \mathbb{F}_2^q$.

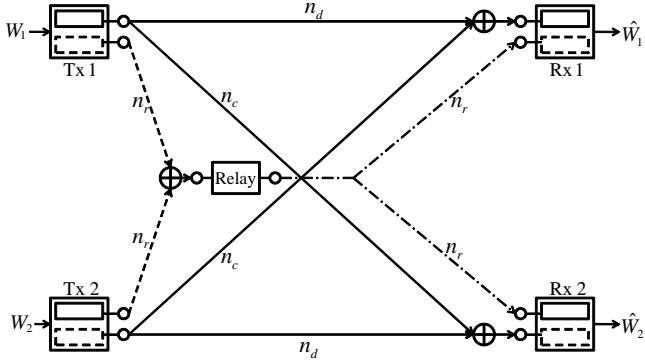

Fig. 2. Deterministic interference channel with an out-of-band half-duplex relay.

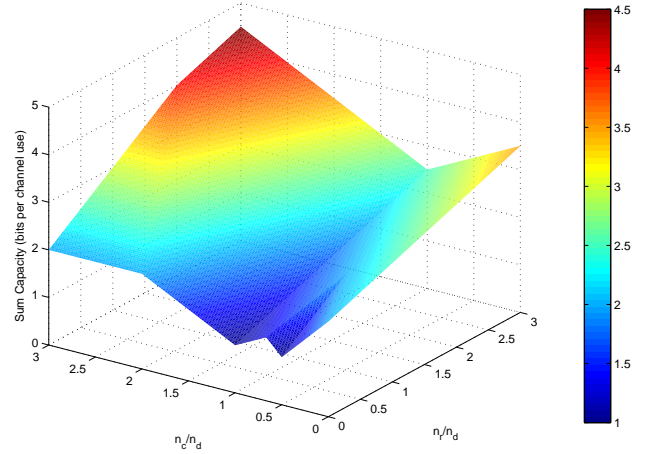

Fig. 3. Sum capacity for the linear deterministic model.

The signal interaction in the deterministic model can be characterized by a series of *add* operations in $\mathbb{F}_2$, and *shift* operations defined by the $q \times q$ matrix

$$\mathbf{S} = \begin{pmatrix} 0 & 0 & 0 & \cdots & 0 \\ 1 & 0 & 0 & \cdots & 0 \\ 0 & 1 & 0 & \cdots & 0 \\ \vdots & \ddots & \ddots & \ddots & \vdots \\ 0 & \cdots & 0 & 1 & 0 \end{pmatrix}. \quad (12)$$

The output of the channel can be characterized as follows: For all $t = \{1, 2, \ldots, m\}$

$$\mathbf{y}_{11,t} = \mathbf{S}^{q-n_d}\mathbf{x}_{11,t} \oplus \mathbf{S}^{q-n_c}\mathbf{x}_{21,t} \quad (13)$$

$$\mathbf{y}_{21,t} = \mathbf{S}^{q-n_c}\mathbf{x}_{11,t} \oplus \mathbf{S}^{q-n_d}\mathbf{x}_{21,t}. \quad (14)$$

For $t = \{1, 2, \ldots, \alpha m\}$

$$\mathbf{y}_{r,t} = \mathbf{S}^{q-n_r}\mathbf{x}_{12,t} \oplus \mathbf{S}^{q-n_r}\mathbf{x}_{22,t}. \quad (15)$$

For $t = \{\alpha m + 1, \ldots, m\}$

$$\mathbf{y}_{12,t} = \mathbf{S}^{q-n_r}\mathbf{x}_{r,t} \quad (16)$$

$$\mathbf{y}_{22,t} = \mathbf{S}^{q-n_r}\mathbf{x}_{r,t}. \quad (17)$$

## III. Sum Rate Optimal Strategies for the Deterministic Symmetric Interference Channel with an Out-of-Band Relay

In this section, we derive outerbounds for the deterministic symmetric IC-OBR using the genie-aided approach, and construct achievable strategies that are sum capacity achieving. Due to the orthogonality between IC and OBRC, we assume for simplicity that $\mathbf{x}_{11}, \mathbf{x}_{21}, \mathbf{y}_{11}, \mathbf{y}_{21}$ are length $\max\{n_d, n_c\}$ vectors, while $\mathbf{x}_{12}, \mathbf{x}_{22}, \mathbf{y}_r, \mathbf{y}_{12}, \mathbf{y}_{22}$ are length $n_r$ vectors. We have the following theorem for sum capacity of this channel.

*Theorem 1:* For the deterministic symmetric interference channel with an out-of-band half-duplex relay, the optimal duplexing factor is $\alpha = 0.5$, and the sum capacity is

$$R_1 + R_2 = \begin{cases} 2n_d + n_r, \\ \qquad \text{when } n_c \geq 2n_d + \frac{1}{2}n_r. \\ n_c + \frac{1}{2}n_r, \\ \qquad \text{when } 2n_d + \frac{1}{2}n_r > n_c \geq n_d. \\ 2n_d - n_c + \frac{1}{2}n_r, \\ \qquad \text{when } n_d > n_c \geq \frac{2}{3}n_d. \\ \min\{2n_c + n_r, 2n_d - n_c + \frac{1}{2}n_r\}, \\ \qquad \text{when } \frac{2}{3}n_d > n_c \geq \frac{1}{2}n_d. \\ \min\{2(n_d - n_c) + n_r, 2n_d - n_c + \frac{1}{2}n_r\}, \\ \qquad \text{when } \frac{1}{2}n_d > n_c. \end{cases}$$

Figure 3 shows how the sum capacity scales with the ratio $\frac{n_c}{n_d}$ and the ratio $\frac{n_r}{n_d}$. We can see that when $\frac{n_r}{n_d}$ is small, the sum capacity has a "W" shape as is the case for the IC [7]. However, as $\frac{n_r}{n_d}$ grows, the "W" curve gradually turns into a "V" curve. This effect is similar to the IC with output feedback, observed in [21], where the sum capacity is shown to have the shape of "V" curve as well. The reason for the improvement in IC with output feedback is that the output feedback provides the sources more information about each other, and thus the sources can utilize the resources in a more efficient manner. For our model, this improvement transpires thanks to the OBRC making the utilization of the available signal resources more efficient, although the sources cannot obtain any information about each other, as we explain later in detail when describing the achievable strategies. We also note that, the sum capacity is unbounded as $\frac{n_c}{n_d} \to \infty$ and $\frac{n_r}{n_d} \to \infty$, whereas the sum capacity of IC saturates as $\frac{n_c}{n_d} \to \infty$.

To prove the theorem, we first derive outerbounds using genie-aided approach. We then show that the outerbounds can be achieved.

*Proposition 1:* The capacity region of the deterministic symmetric interference channel with an out-of-band half-duplex relay is contained in the region $\mathcal{R} = (R_1, R_2)$ specified

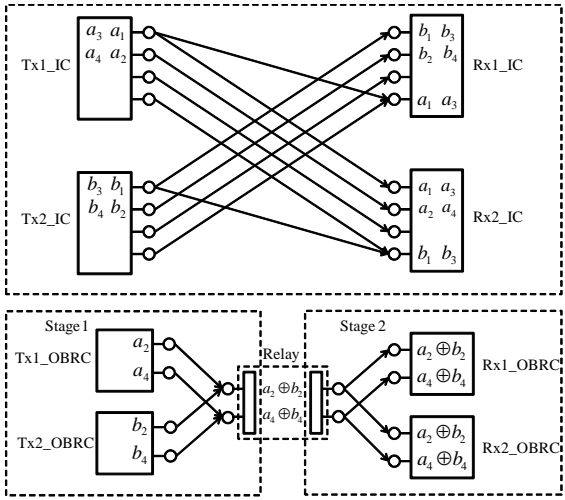

Fig. 4.  Transmission scheme when $n_c = 4$, $n_d = 1$, $n_r = 2$.

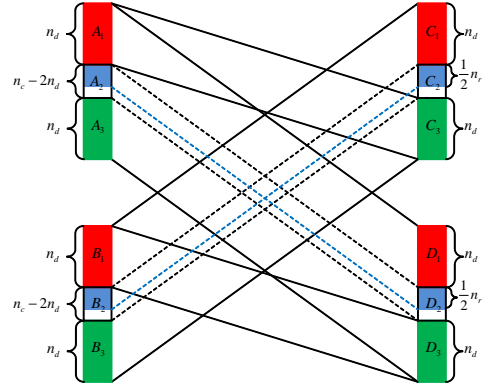

Fig. 5.  Signal interaction for the underlying IC when $n_c \geq 2n_d + \frac{1}{2}n_r$.

by the following rate expressions:

$$R_1 \leq n_d + \frac{1}{2}n_r \tag{18}$$

$$R_2 \leq n_d + \frac{1}{2}n_r \tag{19}$$

$$R_1 + R_2 \leq n_c + \frac{1}{2}n_r, \text{ when } n_c \geq n_d \tag{20}$$

$$R_1 + R_2 \leq \min\{n_r + 2\max\{n_d - n_c, n_c\}, 2n_d - n_c + \frac{1}{2}n_r\},$$
$$\text{when } n_d > n_c. \tag{21}$$

*Proof:* See Appendix A. ∎

Depending on the values of $n_d$, $n_r$ and $n_c$, i.e., the strength of the links, we now construct sum-rate optimal achievable strategies. In particular, for the out-of-band half-duplex relay, we use a two stage transmission scheme with duplexing factor $\alpha = 0.5$, where in the first stage, the relay listens, and in the second stage, the relay transmits. As shown in Appendix A.A this is the optimal duplexing factor for the outerbound, and as shown in the sequel the achievable sum rates with this duplexing factor match the sum rate outer bound, establishing that $\alpha = 0.5$ is sum-capacity optimal. We present our achievable strategies for the following cases:

### A. Case 1: $n_c \geq 2n_d + \frac{1}{2}n_r$

We term this case *extremely strong interference*. To better illustrate the idea of the transmission scheme, we first provide a simple example in Figure 4, where $n_c = 4$, $n_d = 1$, $n_r = 2$. Since interference is extremely strong, all four signal levels at the sources can be received at the interference links, while only the highest signal level can be received at the direct links. The sources transmit information bits $a_1, a_3$ and $b_1, b_3$ to both intended and non-intended destinations using the highest signal levels, and transmit interference signal bits $a_2, a_4$ and $b_2, b_4$ only to the non-intended destinations using the second highest signal level, during two consecutive channel uses. In the first channel use, the sources also transmit signal bits

$a_2, a_4$ and $b_2, b_4$ to the OBR. The OBR receives the sum of the signal bits and then forwards to the destinations in the second channel use. Since destinations have the interference signal bits received from the IC, they can recover the intended information bits from the signals received from the OBR.

Following the above example, we are now ready to illustrate the transmission scheme for the general case. The signal interaction between different signal spaces for the IC is shown in Figure 5, where each part of the signal spaces contains the signal bits in vectors $\mathbf{x}_{i1}$, $\mathbf{y}_{i1}$. For example, the signal space $A_1$ contains the most significant $n_d$ signal bits in vector $\mathbf{x}_{11}$, or the signal level holding information bits $a_1, a_3$ in the above example. Similarly, signal spaces $A_2$ and $A_3$ correspond to the next $n_c - 2n_d$ and $n_d$ signal bits, respectively. Specifically, in the above example, signal spaces $A_2$ and $A_3$ correspond to the second signal level holding bits $a_2, a_4$, and the remaining two empty signal levels, respectively. Without the OBR, each source can only send information bits using the signal spaces which are visible to its intended receiver, e.g., spaces $A_1$ and $B_1$. The other signal spaces, e.g., $A_2, A_3$ and $B_2, B_3$, are unused, since the signals sent from these spaces are only visible to the other destination. With the OBR, part of the unused signal spaces can be utilized to facilitate interference cancelation. Specifically, the sources can use $n_r$ bits of the OBRC *in common* to transmit new information bits, i.e., in the example of the transmission scheme provided in Figure 4, both sources transmit simultaneously to the same signal spaces of the OBR, and the OBR receives the sum of the signal bits from two sources. The OBR simply forwards the received signal bits to the destinations. Since the sources use the signal bits of the OBRC in common, they interfere each other. For the IC, since $n_c \geq 2n_d + \frac{1}{2}n_r$, $\frac{1}{2}n_r$ bits of the spaces $A_2$ and $B_2$ are visible to the other destination without corrupting other signal bits. This can be seen in Figure 5. For two stages, there are $n_r$ bits available from each of the space $A_2$ and $B_2$. The sources can use the spaces $A_2$ and $B_2$ to transmit the signal bits sent through the OBRC as side information to the non-intended destination. These signal bits can be used to cancel the interference in the signal received from the OBRC.

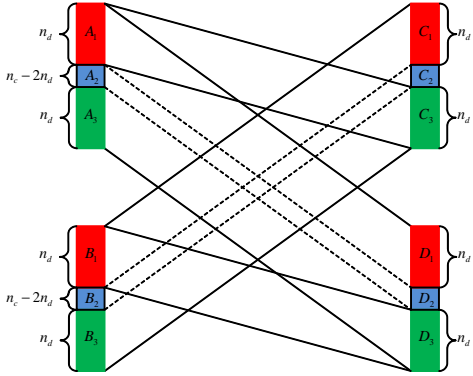

Fig. 6.   Signal interaction for the underlying IC when $2n_d + \frac{1}{2}n_r > n_c \geq 2n_d$.

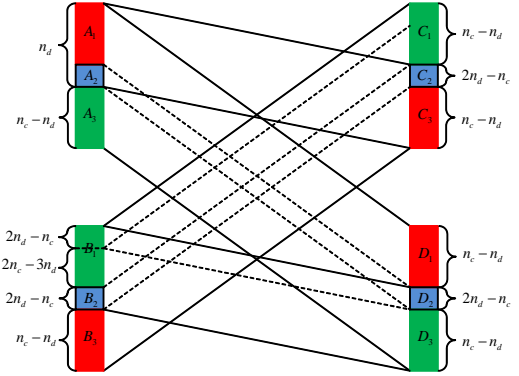

Fig. 7.   Signal interaction for the underlying IC when $2n_d > n_c \geq n_d$.

This scheme achieves the rate pair $(R_1, R_2) = (n_d + \frac{1}{2}n_r, n_d + \frac{1}{2}n_r)$, which is exactly the cut set bound for the individual rates, and thus the capacity region for this scenario is characterized. We can see from here that the channel acts as if there are two independent OBRC helping each source-destination pair, since each pair can achieve a rate of the form $\frac{1}{2}n_r$ through the OBRC, which is the maximum rate one user can achieve using the half-duplex OBR.

### B. Case 2: $2n_d + \frac{1}{2}n_r > n_c \geq 2n_d$

This is the case with *very strong interference*. The signal interaction for this case is shown in Figure 6. Similar to the case in section III-A, without the OBR, each source only transmits information bits using spaces $A_1$ and $B_1$. With the OBR, since $2n_d + \frac{1}{2}n_r > n_c \geq 2n_d$, the sources can use $2(n_c - 2n_d)$ bits of the OBR in common to transmit new information. The rest $n_r - 2(n_c - 2n_d)$ bits of the OBR can be used by one source, or divided between two sources to transmit new information. The $2(n_c - 2n_d)$ common signal bits of the OBR are corrupted by interference. For each stage, the $n_c - 2n_d$ signal bits in spaces $A_2$ and $B_2$ can be used to transmit the signal bits sent through the OBRC to the non-intended destinations as side information, which can help cancel the interference. The sum rate achieved is $n_c + \frac{1}{2}n_r$ bits per channel use, which is exactly the sum capacity of this channel according to the upperbound (20). From the cut set bound for individual rates, we can see that this scheme also achieves the corner points of the capacity region, and thus we can fully characterize the capacity region for this case.

*Remark 1:* So far, we have considered very strong, or extremely strong interference links. The key idea is to let the sources transmit new information bits using the signal spaces of OBRC in common, while the strong interference links can provide some side information for the destination to facilitate interference cancelation. The transmission between

IC and OBRC is *non-separable*, which means that the signals transmitted in the IC and the OBRC are correlated. For the case of extremely strong interference, the signal spaces of the OBRC are limited, compared with the signal spaces available at the interference links. For this case, the sources should use all the signal spaces of the OBRC to transmit new information bits in common such that the side information transmitted through the interference link can be utilized to the fullest extent. Specifically, one bit of the OBRC can help each source transmit one bit, that is, we can trade one bit of the OBRC for the transmission of two information bits. For the case of very strong interference, the OBRC has more signal spaces available than the interference links. The sources can use part of the signal spaces of the OBRC in common to utilize all the side information provided by the interference links, and split the additional signal spaces to transmit new information bits. When the sources split the signal spaces of the OBRC, we trade one bit of the OBRC for the transmission of only one information bit. Thus when using the resources of the OBRC, we should first consider making use of the side information transmitted through the interference links, since this provides the largest payoff. For the following cases when interference is strong or moderate, we will adopt a different approach to construct the optimal transmission strategies.

### C. Case 3: $2n_d > n_c \geq n_d$

This is the case when the interference is *strong*. The signal interaction is shown in Figure 7. Without the OBR, to achieve the sum capacity of the IC, source 1 transmits $n_d$ information bits using the signal space $A_1, A_2$, while source 2 transmits $n_c - n_d$ information bits using all the $2n_d - n_c$ bits in signal space $B_2$ and the lower $2n_c - 3n_d$ bits in signal space $B_1$, as shown in Figure 7. The $2n_d - n_c$ bits in signal space $B_2$ cause interference at the signal space $C_2$ at destination 1. Source 2 uses the higher level $2n_d - n_c$ bits in signal space $B_1$ to transmit another copy of the signal bits in space $B_2$. The higher level $2n_d - n_c$ bits in signal space $B_1$ are visible to destination 1 without corrupting other signals. Destination 1 can thus remove the interference and obtain a clean signal, and the sum capacity of $n_c$ bits can be achieved. Different from the cases in Sections III-A and III-B, with the scheme that achieves sum capacity of the IC, there is no additional signal space available at the sources that does not cause interference at the destinations. Therefore the sources cannot use the signal spaces of the OBRC in common to transmit new information bits. The signal spaces of the OBRC can only be used by one source, or divided between two sources. Since there are $n_r$ bits available at the OBRC, the sum rate achieved in two stages is $n_c + \frac{1}{2} n_r$ bits per channel use. Comparing with the outerbound (20), when $n_c \geq n_d$, this is exactly the sum capacity. The cut set bound for individual rate, the corner points $(n_d + \frac{1}{2} n_r, n_c - n_d)$ and $(n_c - n_d, n_d + \frac{1}{2} n_r,)$ can also be achieved. Thus, the capacity region, for this case, can be characterized as well.

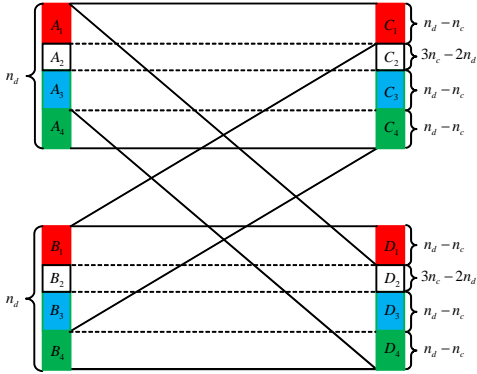

Fig. 8. Signal interaction for the underlying IC for the case $n_d > n_c \geq \frac{2}{3} n_d$.

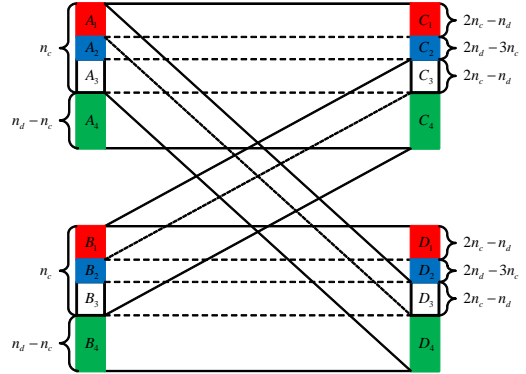

Fig. 9. Signal interaction for the underlying IC when $\frac{2}{3} n_d > n_c \geq \frac{1}{2} n_d$

### D. Case 4: $n_d > n_c \geq \frac{2}{3} n_d$

This is the case with *moderate interference*. The signal interaction for the IC is shown in Figure 8. Without the OBR, it is known that the sum capacity for this case is $R_1 + R_2 = 2n_d - n_c$ [22]. Similar to the case in the previous section (section III-C), for the sum capacity optimal strategy for the IC, there is no additional signal space available at the sources that does not cause interference at the destinations. The signal spaces of the OBR can be used by one source or divided between two sources to transmit $n_r$ new information bits in two stages. The sum rate achieved by this scheme is $R_1 + R_2 = 2n_d - n_c + \frac{1}{2} n_r$ bits per channel use, which matches the outerbound in (21). Thus, the sum capacity is characterized.

*Remark 2:* It is easy to verify that the individual rate $n_d + \frac{1}{2} n_r$ of the cut set bound can be achieved by allowing only one user to use the channel. However, the maximum rate of the other user is 0. The sum rate for this case is less than the sum capacity derived above. The reason is that there may exist another bound of the form $2R_1 + R_2$ which is active in this case. However, it is difficult to obtain an expression for this bound.

*Remark 3:* For the cases described in Sections III-C and III-D, the sources cannot use the signal spaces of the OBRC in common to transmit new information, since no signal space of the IC can be used to cancel the interference in the signal received from the OBRC. The signal spaces of the OBRC can be used by only one source, or divided between two sources to transmit some new independent messages. This shows the optimality of *separate encoding* for the IC and OBRC, i.e., the messages transmitted through IC and OBRC are independent. For the following cases when the interference links are weaker, we will adopt yet another approach to construct the transmission strategy.

### E. Case 5: $\frac{2}{3} n_d > n_c \geq \frac{1}{2} n_d$

This is the case with *weak interference*. The signal interaction in the IC is shown in Figure 9. The signal bits from $A_1, A_2, A_3, B_1, B_2, B_3$ are all common information bits, and the signal bits from $A_4, B_4$ are private information bits. Without the relay, the sum rate optimal transmission strategy for the IC is to use the signal spaces $A_1$ and $B_1$ to transmit common information, which is to be decoded at both destinations, and use the signal spaces $A_4$ and $B_4$ to transmit private information, which is to be decoded at the intended destinations. The condition $\frac{2}{3} n_d > n_c \geq \frac{1}{2} n_d$ guarantees that the bits from signal spaces $A_1, A_4, B_1, B_4$ are aligned at the receivers such that they do not interfere with each other. The remaining signal spaces $A_2, A_3, B_2, B_3$ are left unused, since the information bits transmitted using these signal spaces cause interference at the receivers. We will show that, with the OBR, the interference can be removed, and the sum capacity can be achieved. However, the extent to which we can use the signal spaces $A_2, A_3, B_2, B_3$ depends on the strength of the links in the OBRC, and requires a further classification as follows:

*1)* $n_r \geq 4n_d - 6n_c$: To better illustrate the idea of the achievable strategy, we first provide an example in Figure10. The signal levels holding information bits $a_1$, $a_2$ and $a_4, a_5$ correspond to the spaces $A_1$, $A_2$, and $A_4$ in Figure 9, respec-tively. The empty signal level between $a_2$ and $a_4$ corresponds to the space $A_3$. Similar correspondence holds for the rest source and destinations. The common information bits $a_1, a_6$ and $b_1, b_6$, and the private information bits $a_5, a_{10}$ and $b_5, b_{10}$ are received without any interference at direct links, and do not incur any interference at interference links as well. The common information bits $a_2, a_7$ and $b_2, b_7$ are received without any interference at direct links, but they incur interference to the private information bits $b_4, b_9$ and $a_4, a_9$, respectively. To remove the interference, the sources send the information bits $a_2, a_7$ and $b_2, b_7$ using the same signal levels of OBR. Since $n_r \geq 4n_d - 6n_c$, the sources can divide the rest signal levels of OBR between them to send additional information bits $a_3, a_8$ and $b_3, b_8$. The OBR forwards all the received signal bits to the destinations. Destination 1 then decodes $b_2, b_7$ from the signals received from the OBR. Based on these signal bits, it can decode all the information bits.

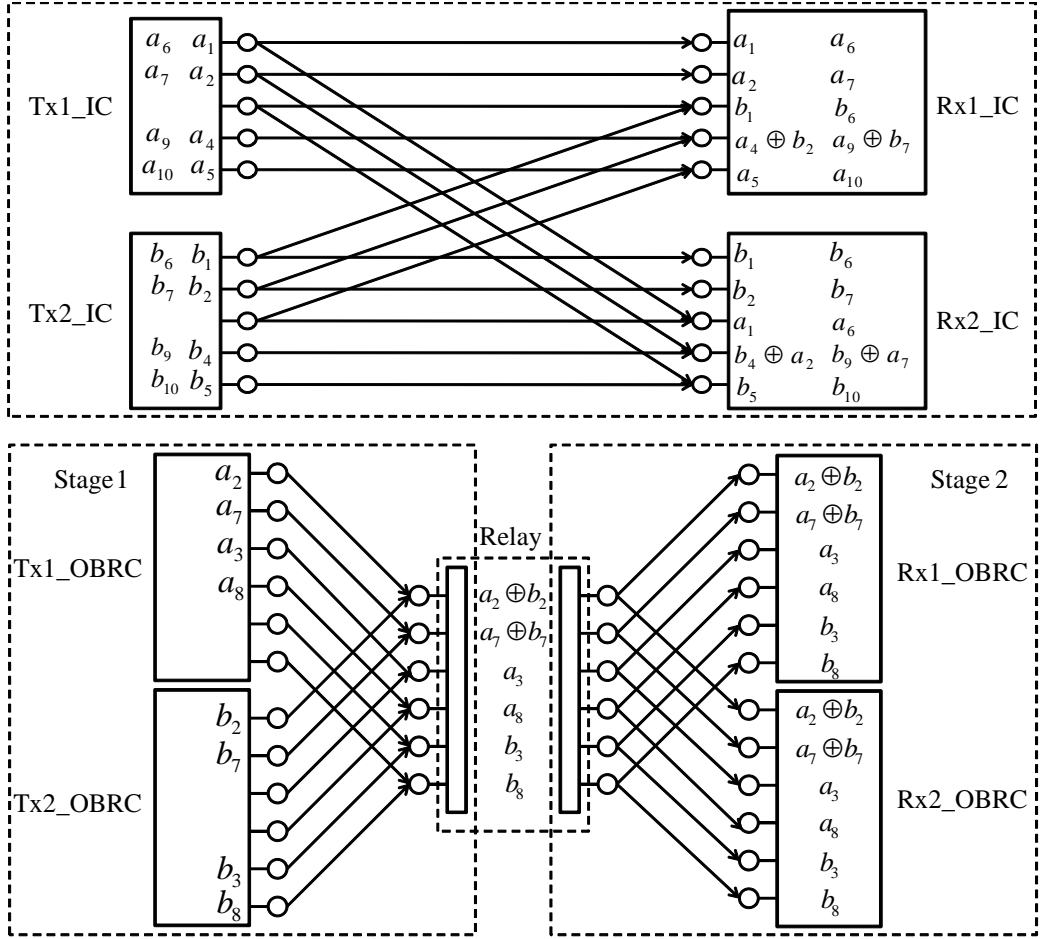

Fig. 10.   Transmission scheme when $n_c = 3$, $n_d = 5$, $n_r = 6$.

Now we are ready to illustrate the strategy for the general case. From the above example, we can see that the difference between the strategies for IC-OBR and IC is that for IC-OBR, the sources can use all signal bits in spaces $A_2, B_2$, in addition to spaces $A_1, B_1, A_4, B_4$ to transmit new information in both stages through the IC. Note that the signal bits transmitted from spaces $A_1, A_2$ and $B_1, B_2$ can be decoded directly at the intended destinations since they are not corrupted by interference. However, $2n_d - 3n_c$ signal bits received at spaces $C_4$ and $D_4$ are corrupted by interference for each stage. The sources use $4n_d - 6n_c$ bits of the OBRC in common to transmit the signal bits in $A_2$ and $B_2$, and the rest $n_r - 4n_d + 6n_c$ signal bits of the OBRC can be used by one source or shared between two sources to transmit additional new information. The relay simply forwards all the information bits to the destinations. At the destinations, the $4n_d - 6n_c$ bits received from the OBRC carry the modulo sum of information bits from spaces $A_2$ and $B_2$. Since each destination knows the signal bits from one of the spaces, it can recover the signal bits from the other space. Therefore the interference bits in spaces $C_4$ and $D_4$ can be removed. The sum rate can be achieved is $2n_d - n_c + \frac{1}{2}n_r$, which matches the outerbound (21).

*2) $4n_d - 6n_c > n_r$:* For this case, since the resources of the relay are limited, the sources can only use $\frac{1}{2}n_r$ bits of the spaces $A_2$ and $B_2$ to transmit signals into the IC for each stage, in addition to the spaces $A_1, A_4$ and $B_1, B_4$. All the signal bits in the OBRC are used in common by two sources to transmit the signal bits from spaces $A_2$ and $B_2$ in two stages. At destination 1, the decoder first decodes the signal bits transmitted from space $A_1$, and part of the space $A_2$. It can then recover the interference signal bits sent from space $B_2$ utilizing the OBRC. With all the interfering signal bits, it can decode all the information bits. The sum rate achieved is $n_r + 2n_c$, which matches the outerbound (21).

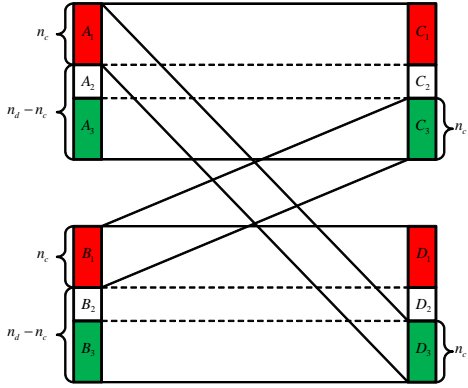

Fig. 11. Signal interaction for the underlying IC when $\frac{1}{2}n_d > n_c$.

*Remark 4:* Note that for weak interference, we only utilize the common information bits from $A_2$ and $B_2$ to transmit new information bits, but the signal spaces $A_3$ and $B_3$ are left unused. The reason is that the signal bits from $A_2$ and $B_2$ only cause interference at $D_4$ and $C_4$, respectively, but they are not interfered by other signal bits. However, the signal bits from $A_3$ and $B_3$ not only cause interference at the other destination, but they are also interfered by the other source. To recover one bit from $A_2$ and one bit at the corresponding level from $B_2$, we only need one bit from the OBRC, that is, we trade one bit of the OBRC for the transmission of two information bits. However, to cancel the interference caused by using one bit from $A_3$ and one bit from the corresponding level of $B_3$, we need two bits from the OBRC, i.e., we only trade one bit of the OBRC for the transmission of one information bit, which is the same as the case when the signal spaces of the OBRC are used by only one source, or divided between two sources, to transmit new information. In addition, using the spaces $A_3$ and $B_3$ makes the signal interaction more complicated, and requires a more involved achievable strategy.

### F. Case 6: $\frac{1}{2}n_d > n_c$

This is the case with *very weak interference*. The signal interaction for the IC is shown in Figure 11. Without the OBR, the optimal transmission scheme is to transmit "private" information, i.e., to transmit information using signal spaces $A_2$, $A_3$ and $B_2$, $B_3$, since these signal bits are invisible to the other receiver. The signal spaces $A_1$ and $B_1$ are left unused, since the signal bits from these spaces are common information bits, and they cause interference at the destinations. With the OBR, the signal spaces $A_1$ and $B_1$ can be utilized to transmit additional information bits, and the resulting interference can be removed. Similar to the strategy described in section III-E, the extent to which we can use the signal spaces $A_1$ and $B_1$ depends on the strength of the links in the OBRC. Thus we consider the following subcases.

*1) $n_r \geq 2n_c$:* For this case, the sources use all the signal spaces to transmit information through the IC. In the OBRC, each source simultaneously transmits the $2n_c$ signal bits in spaces $A_1$ and $B_1$ using $2n_c$ signal bits of the OBRC. Each signal bit received at the OBR is the sum in $\mathbb{F}_2$ of the corresponding signal bits from two sources. The remaining $n_r - 2n_c$ bits of the OBR can be used by one source, or divided between two sources to transmit new information bits. At destination 1, the decoder first decodes the signal bits sent from space $A_1$. It can then recover the interfering signal bits from $B_1$ using the signal obtained from the OBRC. With all the interference signal bits, it can decode all the intended information bits. The sum rate achieved is $2n_d - n_c + \frac{1}{2}n_r$ bits per channel use, which coincides with the upperbound (21).

*2) $2n_c > n_r$:* For this case, since the signal spaces at the OBRC are limited, the sources can transmit their information bits using all signal spaces $A_2$, $A_3$ and $B_2$, $B_3$, and $\frac{1}{2}n_r$ bits of spaces $A_1$ and $B_1$ for each stage. All the signal bits in the OBRC are used in common by two sources to transmit the signal bits from spaces $A_1$ and $B_1$. The sum rate achieved for this case is $n_r + 2n_d - 2n_c$ bits per channel use, which matches the upperbound (21).

*Remark 5:* From the transmission scheme described in Section III-E and III-F, we can see that the signal bits transmitted through IC and OBRC are correlated, and thus for these two cases, the optimal strategy is to use IC and OBRC in the *non-separable* fashion.

*Remark 6:* For the channel settings discussed in Section III-E.2 and III-F.2, the OBRC cannot help the sources to transmit new information bits. It can only facilitate interference cancelation. However, for the channel settings discussed in Section III-E.1 and III-F.1, the OBRC can help the sources to transmit new information bits in addition to facilitate interference cancelation, since the OBRC has more resources to be utilized. When the OBRC is used for interference cancelation, one bit of the OBRC can help each source to transmit one information bit, which means we trade one bit of the OBRC for two information bits. The optimality of our achievable strategy shows that when using the resources of the OBRC under weak and very weak interference, we should first consider using the OBRC to facilitate canceling the interference caused by transmitting additional common information bits, since this provides the largest payoff.

### G. Summary for the Deterministic Model

So far, we have characterized the sum capacity of the linear deterministic IC-OBR. We have shown that the OBRC can make the resource utilization more efficient. In the following remarks, we provide a brief summary of the design insights for optimal achievable strategies obtained from the deterministic model, in order to make connections with the Gaussian model.

*Remark 7:* **Extremely strong interference:** $n_c \geq 2n_d + \frac{1}{2}n_r$. The optimal strategy is to use interference link to transmit side information to the destination. All the signal spaces of the relay are designated to utilize the side information from the interference links. Sources use the signal spaces of the relay in common. Non-separable encoding between IC and OBRC is optimal.

*Remark 8:* **Very strong interference:** $2n_d + \frac{1}{2}n_r > n_c \geq 2n_d$. The optimal strategy is similar to the extremely strong interference case. The difference is that the relay has additional signal spaces to help the sources transmit some new messages. Sources use part of the signal spaces of the relay in common. Non-separable encoding between IC and OBRC is optimal.

*Remark 9:* **Strong interference:** $2n_d > n_c \geq n_d$. Separate encoding between IC and OBRC is optimal. Destinations use successive interference cancellation for the IC, and the signal spaces of the relay are divided between the sources.

*Remark 10:* **Moderate interference:** $n_d > n_c \geq \frac{2}{3}n_d$. Separate encoding between IC and OBRC is optimal. Han-Kobayashi strategy is employed for the IC and the signal spaces of the relay are divided between the sources.

*Remark 11:* **Weak interference:** $\frac{2}{3}n_d > n_c \geq \frac{1}{2}n_d$. The optimal strategy is to let the sources use a modified version of Han-Kobayashi strategy to transmit some new common information bits, and the relay is used to cancel the additional interference caused by the new common information bits. Depending on the strength of the relay links, sources can use all of the signal spaces, or part of the signal spaces of the relay in common. Non-separable encoding between IC and OBRC is optimal.

*Remark 12:* **Very weak interference:** $\frac{1}{2}n_d > n_c$. The optimal strategy is to use Han-Kobayashi strategy to transmit both common and private information bits. The interference caused by common information bits can be canceled using the relay. Depending on the strength of the relay links, sources can use all of the signal spaces, or part of the signal spaces of the relay in common. Non-separable encoding between IC and OBRC is optimal.

Now, based on the insights obtained from the deterministic model, we are ready to study the Gaussian model.

## IV. The Symmetric Gaussian Interference Channel with an Out-of-Band Relay

In this section, we consider the Gaussian interference channel with an out-of-band relay with duplexing factor 0.5. For the Gaussian channel, it is not clear whether the duplexing factor 0.5 is optimal. However, as we will show in the sequel, the gap between the outerbounds with optimal duplexing factor and the outerbounds with duplexing factor 0.5 is small. Therefore any constant gap result with duplexing factor 0.5 implies constant gap result with the optimal duplexing factor. In addition, our main goal is to assess the impact of interference and relaying strategies on this model. With the fixed duplexing factor, we are able to illustrate the interaction between interference and OBR in a clearer fashion.

Recall that for the symmetric channel, rate points which achieve the sum capacity also achieve the symmetric capacity.

We thus investigate the sum capacity of the Gaussian channel. We first derive outerbounds for the Gaussian IC-OBR.

*Proposition 2:* When the interference links are stronger than the direct links, i.e., $h_c^2 \geq h_d^2$, the following expressions provide sum rate upperbounds:

$$C_{sum,1} \leq \log(1 + h_d^2) + \min\left\{\alpha\log(1 + h_c^2),\right.$$
$$\left.(1 - \alpha)\log(1 + h_r^2)\right\} \quad (22)$$

$$C_{sum,2} \leq \frac{1}{2}\log(1 + h_c^2 + h_d^2) + \min\left\{\frac{\alpha}{2}\log(1 + 2h_r^2),\right.$$
$$\left.\frac{1 - \alpha}{2}\log(1 + h_r^2)\right\}. \quad (23)$$

When the interference links are weaker than the direct links, the following expressions provide sum rate upperbounds:

$$C_{sum,3} \leq \log\left(1 + h_c^2 + \frac{h_d^2}{1 + h_c^2}\right) + \min\left\{\alpha\log(1 + 2h_r^2),\right.$$
$$\left.(1 - \alpha)\log(1 + h_r^2)\right\} \quad (24)$$

$$C_{sum,4} \leq \frac{1}{2}\log\left(1 + h_d^2\right) + \min\left\{\frac{\alpha}{2}\log(1 + 2h_r^2),\right.$$
$$\left.\frac{1 - \alpha}{2}\log(1 + h_r^2)\right\}. \quad (25)$$

*Proof:* See Appendix B. ∎

*Remark 13:* Note that the bound $C_{sum,4}$ is a special form of the sum rate outerbound derived in a recent work [17]. For the bound $C_{sum,2}$ and $C_{sum,4}$, we utilized the symmetry of the channel, that is, the channel outputs at the OBRC have the same statistics at the receivers.

*Proposition 3:* The sum rate outerbound evaluated at $\alpha = 0.5$ at most has a finite gap of 0.25 bits with the sum rate outerbound evaluated at optimum $\alpha^*$.

*Proof:* We assume $h_r \geq 1$, since otherwise the terms associated with $h_r$ are less than 1. Denote the sum rate outerbound in *Proposition 2* as

$$C^{opt} = \max_\alpha \min\left\{C_{sum,1}, C_{sum,2}, C_{sum,3}, C_{sum,4}\right\}. \quad (26)$$

Note that $C^{opt} \leq \max_\alpha C_{sum,i}$, $i = 1, 2, 3, 4$. We also denote the sum rate outerbounds evaluated at $\alpha = 0.5$ as $C_{sum,i}^{0.5}$. It is easy to see that

$$\max_\alpha C_{sum,1} - C_{sum,1}^{0.5} = 0. \quad (27)$$

We also have

$$\max_\alpha C_{sum,2} - C_{sum,2}^{0.5} \quad (28)$$

$$= \frac{\log(1 + 2h_r^2)}{2\left(\log(1 + 2h_r^2) + \log(1 + h_r^2)\right)}\log(1 + h_r^2)$$
$$- \frac{1}{4}\log(1 + h_r^2) \quad (29)$$

$$= \frac{\left(\log(1 + 2h_r^2) - \log(1 + h_r^2)\right)}{4\left(\log(1 + 2h_r^2) + \log(1 + h_r^2)\right)}\log(1 + h_r^2) \quad (30)$$

$$= \frac{\left(\log\left(\frac{1 + 2h_r^2}{1 + h_r^2}\right)\right)}{4\left(\frac{\log(1 + 2h_r^2)}{\log(1 + h_r^2)} + 1\right)} \quad (31)$$

$$\leq \frac{1}{8}. \tag{32}$$

$C_{sum,3}^{0.5}$ and $C_{sum,4}^{0.5}$ can be evaluated in similar fashion, where the gap between $C_{sum,3}^{0.5}$ and $\max_\alpha C_{sum,3}$ is 0.25. ∎

Therefore, we use outerbounds evaluated at $\alpha = 0.5$ in the sequel. Note that for simplicity, we denote $C_{sum,i}^{0.5}$ as $C_{sum,i}$.

We are now ready to show that we can achieve within constant gap of the above outerbounds. In the remainder of the paper, we construct achievable strategies based on the insights obtained from the deterministic model and calculate the achievable rates. Depending on the relative strength between the interference link and the direct link, we study both weak and strong interference regimes. For each case, we first propose an achievable rate based on strategies which are extensions of the ones used for IC, and then identify channel settings where the constant gap result can be established. For other channel settings, we design new achievable strategies to establish the constant gap results. We focus our study on the case when $h_d^2 \geq 1$, $h_c^2 \geq 1$, since this is of our primary interest. For the cases when $h_d^2 < 1$, $h_c^2 < 1$, we can extend the strategy by treating the signals come from weak links as noise. We present our results as follows:

### A. When the interference link is stronger than the direct link: $h_c^2 \geq h_d^2$

*Proposition 4:* When the interference link is stronger than the direct link, the following sum rate is achievable

$$R_{sum} = \min\{\log(1+h_c^2), \frac{1}{2}\log(1+h_d^2+h_c^2)\} + \frac{1}{4}\log(1+h_r^2). \tag{33}$$

*Proof:* To show the achievability of this rate, we propose a strategy which is a simple extension of the strategy used in the IC. Each source splits the message $W_i$ into two parts, $W_{iD}$ and $W_{iR}$, where we send $W_{iD}$ through the IC, while $W_{iR}$ through the OBRC. For the IC, the destinations decode both messages. For the OBRC, the relay treats the signal received from the incoming bands as a MAC. It decodes both messages, encodes the messages with equal power, and sends the messages to the destinations using the outgoing bands. It is easy to verify that the sum rate can be achieved. ∎

We now evaluate the rate (33) for the following cases:

*1) $h_c^2 \leq h_d^2 + h_d^4$:* Under this condition, the rate expression reduces to

$$R_{sum} = \frac{1}{2}\log(1+h_d^2+h_c^2) + \frac{1}{4}\log(1+h_r^2) \tag{34}$$

which matches the outerbound $C_{sum,2}$ in (23).

*Relation to the deterministic model:* The condition $h_d^2 \leq h_c^2 \leq h_d^2 + h_d^4$ corresponds to $2n_d > n_c \geq n_d$, i.e., **strong interference**, for the deterministic model. Recalling the summary provided in *Remark 9*, we notice that the achievable strategy for the Gaussian model complies with the insights obtained from the deterministic model, i.e., separate encoding between the IC and OBRC is optimal.

*2) $h_c^2 > h_d^2 + h_d^4$:* For this case, the achievable rate (33) reduces to

$$R_{sum} = \log(1+h_d^2) + \frac{1}{4}\log(1+h_r^2), \tag{35}$$

which has unbounded gap with the outerbounds. To establish a constant gap result, we need to design new achievable strategy to improve the rate.

*Relation to the deterministic model:* The condition $h_c^2 > h_d^2 + h_d^4$ corresponds to the case $n_c \geq 2n_d$, i.e., very/extremely strong interference, for the deterministic model. Recalling from the summary in *Remark 7* and *Remark 8*, we notice that the achievable strategy for the deterministic model motivates us to let the sources utilize the very strong interference links to transmit additional messages to the non-intended receivers, and align them at the noise level at the direct links. In addition, the sources also let the OBR forward the sum of these messages to the destinations, where the intended messages can be decoded with the side information from the interference links.

With the insights obtained from the deterministic model, we will now demonstrate that the following rate is achievable for $h_c^2 > h_d^2 + h_d^4$.

*Proposition 5:* When interference is **extremely strong**, i.e.,

$$\left(\frac{1}{4}\log\left(\frac{1}{2}+h_r^2\right)\right)^+ \leq \frac{1}{2}\log\left(1+\frac{h_c^2}{h_d^4+h_d^2}\right), \tag{36}$$

the rate

$$R_{sum} = \log\left(1+\frac{h_d^2-1}{2}\right) + \left(\frac{1}{2}\log\left(\frac{1}{2}+h_r^2\right)\right)^+ \tag{37}$$

is achievable. Otherwise, when interference is **very strong**, i.e., condition (36) does not hold, the rate

$$R_{sum} = \log\left(1+\frac{h_d^2-1}{2}\right) + \log\left(1+\frac{h_c^2}{h_d^4+h_d^2}\right) + \frac{1}{4}\log\left(\frac{1+h_r^2}{1+h_r^2\gamma^2}\right) \tag{38}$$

is achievable, where the parameter $\gamma$ is chosen such that the following condition holds

$$\frac{1}{4}\log\left(\frac{1}{2}+h_r^2\gamma^2\right) = \frac{1}{2}\log\left(1+\frac{h_c^2}{h_d^4+h_d^2}\right). \tag{39}$$

Moreover, the above sum rates (37) and (38) have constant gap with the outerbounds.

*Proof:* To apply the insights obtained from deterministic model to the Gaussian channel, we consider using lattice codes in the OBRC and Gaussian code in the IC. Each source splits the message $W_i$ into $W_{iD}$ and $W_{iC}$ with rates $R_{iD}$ and $R_{iC}$ respectively, where $W_{iD}$ is to be decoded from the direct link, and $W_{iC}$ is to be decoded from the interference link. The sources then encode $W_{iD}$, $W_{iC}$ into $U_i^m$, $V_i^m$ respectively, where $U_i$, $V_i$ are independent unit variance Gaussian random variables. The signals transmitted into the IC are

$$X_i = \beta U_i + \sqrt{1-\beta^2}V_i. \tag{40}$$

We further choose a pair of nested lattice codes $\Lambda \subset \Lambda_c \subset \mathbb{R}^m$ with nesting ratio $R_{iC}$, such that the coarse lattice $\Lambda$ is Rogers-good and Poltyrev-good [26], and the fine lattice $\Lambda_c$ is Poltyrev-good. Moreover, we choose the coarse lattice such that $\sigma^2(\Lambda) = 1$. The codewords are the fine lattice points that are within the fundamental Voronoi region of the coarse lattice. Source $i$ maps the message $W_{iC}$ into a lattice point $t_i^m \in \Lambda_c \bigcap \mathcal{V}(\Lambda)$, and transmits

$$X_{iR}^m = (t_i^m + D_i^m) \mod \Lambda, \tag{41}$$

where $D_i^m \sim \text{Unif}(\mathcal{V}(\Lambda))$ is the dither. It can be shown that $X_{iR}^m$ satisfies the power constraint and is independent of $t_i^m$ [27].

To guarantee that the messages $W_{iC}$ arrive at the noise level at the direct links, we set $\beta^2 = \frac{h_d^2 - 1}{h_d^2}$. Therefore, the received signal at receiver 1 is given by

$$Y_1 = \sqrt{h_d^2 - 1}U_1 + \frac{h_c}{h_d}\sqrt{h_d^2 - 1}U_2 + \frac{h_c}{h_d}V_2 + Z_1 + V_1. \tag{42}$$

The message $W_{2D}$ is decoded first. Successful decoding requires

$$R_{2D} \leq \frac{1}{2}\log\left(1 + \frac{h_c^2(h_d^2 - 1)}{h_d^4 + h_d^2 + h_c^2}\right). \tag{43}$$

The decoder then tries to recover the message $W_{2C}$. To guarantee vanishing error probability, we need

$$R_{2C} \leq \frac{1}{2}\log\left(1 + \frac{h_c^2}{h_d^4 + h_d^2}\right). \tag{44}$$

The message $W_{1D}$ is decoded last by treating $V_1$ as noise. The rate constraint for this step is

$$R_{1D} \leq \frac{1}{2}\log\left(1 + \frac{h_d^2 - 1}{2}\right). \tag{45}$$

*Remark 14:* Note that from (44), we can see that this rate is positive if and only if the very strong interference condition $h_c^2 > h_d^2 + h_d^4$ is satisfied. Also, from (45), we can see that since we let the message $W_{iC}$ arrives at noise level at the intended destination, there is only 0.5 bits rate loss for the message $W_{iD}$ caused by sending the side information $W_{iC}$ to the non-intended destinations, compared with (33).

For the OBRC, the relay first decodes the modulo sum of the transmitted lattice points from the sources. This is possible if

$$R_{1C} = R_{2C} \leq \left(\frac{1}{4}\log\left(\frac{1}{2} + h_r^2\right)\right)^+ \tag{46}$$

where $(x)^+ = \max\{x, 0\}$[1]. The relay then transmits the modulo sum of the two lattice points, which is a lattice point, to the destinations using the outgoing bands. To guarantee successful decoding at the destination, we need

$$R_{1C} = R_{2C} \leq \frac{1}{4}\log\left(1 + h_r^2\right). \tag{47}$$

Since destination 1 knows $W_{2C}$, it can recover $W_{1C}$ from the signal received from the OBRC. The decoding process at destination 2 is the same as at destination 1. Note that $R_1 = R_{1C} + R_{1D}$, $R_2 = R_{2C} + R_{2D}$. It is easy to verify that when (36) holds, the rate (37) is achievable.

It can further be readily verified that the gap between this rate and $C_{sum,1}$ (22) is $\frac{1}{2}\log 6 = 1.2925$ bits.

*Relation to the deterministic model:* The condition (36) corresponds to the case $n_c \geq 2n_d + \frac{1}{2}n_r$ in the deterministic model, i.e., the interference is extremely strong. We can also interpret this as the resources in the OBRC are limited, and thus the OBRC can only be used to utilize the side information provided by the interference links.

When $2n_d + \frac{1}{2}n_r > n_c \geq 2n_d$, or the condition (36) does not hold, in addition to fully utilizing the side information provided by the interference links, we can use the OBRC to transmit independent new information. The detailed strategy is described next.

Based on the achievable strategy described above, the sources encode additional messages $W_{iR}$ into $U_{iR}$, and transmit $X'_{iR} = \gamma X_{iR} + \sqrt{1 - \gamma^2}U_{iR}$ into the OBRC. The relay first decodes $U_{1R}$ and $U_{2R}$ by treating $X_{iR}$ as noise. To guarantee vanishing error probability, we need a multiple access channel (MAC) type constraint at the relay, where the sum rate constraint is

$$R_{1R} + R_{2R} \leq \frac{1}{4}\log\left(1 + \frac{2h_r^2(1 - \gamma^2)}{1 + 2h_r^2\gamma^2}\right). \tag{48}$$

The relay then subtracts $U_{iR}$ from the received signal, and decodes the modulo sum of the lattice points representing $W_{1C}$ and $W_{2C}$. This requires

$$R_{1C} = R_{2C} \leq \frac{1}{4}\log\left(\frac{1}{2} + h_c^2\gamma^2\right). \tag{49}$$

We denote the modulo sum of these lattice points by $T_R$. The relay transmits

$$X_R = \gamma T_R + \sqrt{\frac{1 - \gamma^2}{2}}(U_{1R} + U_{2R}). \tag{50}$$

The destinations follow the same decoding order as the relay, i.e., they first decode $U_{1R}$, $U_{2R}$ as a MAC, and then decode $T_R$. To guarantee low error probability, we need

$$R_{1R} + R_{2R} \leq \frac{1}{4}\log\left(1 + \frac{h_r^2(1 - \gamma^2)}{1 + h_r^2\gamma^2}\right) \tag{51}$$

and

$$R_{1C} = R_{2C} \leq \frac{1}{4}\log\left(1 + h_r^2\gamma^2\right). \tag{52}$$

We set the parameter $\gamma$ such that the rate constraints of the message $W_{iC}$ are the same for both the IC and the OBRC, which gives us condition (39). Note that $R_1 = R_{1C} + R_{1D} + R_{1R}$, $R_2 = R_{2C} + R_{2D} + R_{2R}$. The achievability of rate (38) can be established.

It can be shown that the gap with the bound $C_{sum,2}$ (23) is thus at most 1.25 bits. For details, see Appendix C. ∎

*Remark 15:* Note that the above strategy, in which we use the OBRC to transmit new information in addition to cooperation with the IC, is used repeatedly in the paper. Since the steps are similar, we will refrain from describing the scheme again in detail in the sequel.

### B. When the interference link is weaker than the direct link: $h_c^2 < h_d^2$

For the IC, Han-Kobayashi scheme [4] yields the largest known achievable rate region for this range of channel parameters, where the messages are splitted into common and private parts. We first present an achievable rate using a simple extension of the Han-Kobayashi scheme.

---

[1] The prelog factor $\frac{1}{4}$ is due to the duplexing factor 0.5.

*Proposition 6:* The following rate is achievable for the IC-OBR using Han-Kobayashi scheme:

$$R_{sum} = \log\left(1 + \frac{h_d^2}{2h_c^2}\right) + \frac{1}{4}\log\left(1 + h_r^2\right)$$
$$+ \min\left\{\frac{1}{2}\log\left(1 + \frac{(h_d^2 + h_c^2)(h_c^2 - 1)}{2h_c^2 + h_d^2}\right),\right.$$
$$\left.\log\left(1 + \frac{h_c^2(h_c^2 - 1)}{2h_c^2 + h_d^2}\right)\right\}. \quad (53)$$

*Proof:* We first split the message $W_i$ into $W_{ic}$, $W_{ip}$ and $W_{iR}$, and then encode the message $W_{ic}$, $W_{ip}$ and $W_{iR}$ into $U_i^m$, $V_i^m$ and $V_{iR}^m$ respectively, where $U_i, V_i, V_{iR} \sim \mathcal{N}(0,1)$. $W_{ic}$ and $W_{ip}$ are common and private messages to be sent through the IC, while $W_{iR}$ is the message to be sent through the OBRC. In light of the result in [7], we let the signals carrying the private information arrive at the noise level at the interference link. The signals transmitted from the sources are

$$X_i = \beta U_i + \sqrt{1 - \beta^2}V_i \quad (54)$$
$$X_{iR} = V_{iR} \quad (55)$$

where $\beta^2 = \frac{h_c^2 - 1}{h_c^2}$. The decoders follow the decoding rule used in [7] for the IC. For the OBRC, the relay treats the signal received from its incoming bands as a MAC, and decodes both messages. It then equally splits its power to transmit both messages using its outgoing bands. The sum rate achieved is (53). ∎

We now evaluate the achievable sum rate (53) for the following two cases:

*1) $h_d^2(h_d^2 + h_c^2) \leq h_c^4(h_c^2 + 1)$:* Under this condition, the above sum rate reduces to

$$R_{sum} = \frac{1}{2}\log\left(1 + h_d^2 + h_c^2\right) + \frac{1}{2}\log\left(2 + \frac{h_d^2}{h_c^2}\right)$$
$$+ \frac{1}{4}\log\left(1 + h_r^2\right) - 1. \quad (56)$$

It is easy to verify that the gap between this rate and the upperbound $C_{sum,4}$ in (25) is 1 bit.

*Relation to the deterministic model:* The conditions $h_d^2(h_d^2 + h_c^2) \leq h_c^4(h_c^2 + 1)$ and $h_c^2 < h_d^2$ correspond to the case $n_d > n_c \geq \frac{2}{3}n_d$, i.e., **moderate interference**, in the deterministic model, where it is optimal for the sources to use separate encoding for the IC and OBRC, as summarized in *Remark 10*. Therefore the insights obtained from the deterministic model comply with the results for the Gaussian model.

*2) $h_d^2(h_d^2 + h_c^2) > h_c^4(h_c^2 + 1)$:* Under this condition, the sum rate (53) has unbounded gap with the outerbounds. To establish the constant gap result, we need to design new achievable strategies to improve the rates.

*Relation to the deterministic model:* This condition corresponds to the case of weak or very weak interference, i.e., $\frac{2}{3}n_d > n_c$, in the deterministic model. For these two cases, the summary for the deterministic model in *Remark 11* and *Remark 12* suggests us to utilize the relay to decode the common messages in the most efficient manner, i.e., the sources should use the signal spaces of the relay in common as much as possible. In the sequel, we elaborate on the detailed achievable strategies for both cases.

*Proposition 7:* For **weak interference**: $h_d^2(h_d^2 + h_c^2) > h_c^4(h_c^2 + 1) \geq (h_d^2 + h_c^2)(h_c^2 + 1)$, when the following condition holds

$$\frac{1}{2}\log\left(1 + \frac{h_d^4}{h_c^6 + h_c^4 + h_d^2 h_c^2}\right) \geq \left(\frac{1}{4}\log\left(\frac{1}{2} + h_r^2\right)\right)^+, \quad (57)$$

the following sum rate is achievable

$$R_{sum} \leq \log\left(1 + \frac{h_d^2}{2h_c^2}\right) + \log\left(1 + \frac{h_c^4 - h_c^2 - h_d^2}{2h_d^2 + 2h_c^2}\right)$$
$$+ \left(\frac{1}{2}\log\left(\frac{1}{2} + h_r^2\right)\right)^+. \quad (58)$$

Otherwise, the following sum rate is achievable

$$R_{sum} \leq \log\left(1 + \frac{h_d^2}{2h_c^2}\right) + \log\left(1 + \frac{h_c^4 - h_c^2 - h_d^2}{2h_d^2 + 2h_c^2}\right)$$
$$+ \log\left(1 + \frac{h_d^4}{h_c^6 + h_c^4 + h_d^2 h_c^2}\right) + \frac{1}{4}\log\left(\frac{1 + h_r^2}{1 + h_r^2\gamma^2}\right), \quad (59)$$

where the parameter $\gamma$ is chosen such that the following condition holds

$$\frac{1}{2}\log\left(1 + \frac{h_d^4}{h_c^6 + h_c^4 + h_d^2 h_c^2}\right) = \frac{1}{4}\log\left(\frac{1}{2} + h_r^2\gamma^2\right). \quad (60)$$

Moreover, the above sum rates (58) and (59) have constant gap with the outerbounds.

*Proof:* From the insights obtained from the deterministic model, i.e., Figure 9 and the strategy described in Section III-E, we can see that the sources should split their messages into three parts, corresponding to the spaces $A_1$, $A_2$ and $A_4$ ($B_1$, $B_2$ and $B_4$). The signal transmitted from $A_4$ and $B_4$ are aligned at the noise level at interference links. At receiver 1 (2), the signal transmitted from $B_2$ ($A_2$) are aligned at the same level as the signal transmitted from $A_4$ ($B_4$).

Based on this insight, we split the message $W_i$ into $W_{ic}$ and $W_{ip}$. We further split the common message $W_{ic}$ into $W_{ica}$ and $W_{icb}$, and encode $W_{ica}$, $W_{icb}$ and $W_{ip}$ into $U_{ia}^m$, $U_{ib}^m$, $V_i^m$ respectively, where $U_{ia}, U_{ib}, V_i \sim \mathcal{N}(0,1)$. The sources thus transmit the following signal through the IC:

$$X_i = \beta(\theta U_{ia} + \sqrt{1 - \theta^2}U_{ib}) + \sqrt{1 - \beta^2}V_i. \quad (61)$$

The signal received at destination 1 is

$$Y_1 = \frac{h_d}{h_c}\sqrt{h_c^2 - 1}(\theta U_{1a} + \sqrt{1 - \theta^2}U_{1b}) + \frac{h_d}{h_c}V_1$$
$$+ \sqrt{h_c^2 - 1}(\theta U_{2a} + \sqrt{1 - \theta^2}U_{2b}) + Z_1 + V_2. \quad (62)$$

The optimal achievable scheme for the deterministic model implies that we should choose the parameter $\theta$ such that $U_{2b}$ and $V_1$ are aligned at the same level. We have

$$1 - \theta^2 = \frac{h_d^2}{h_c^2(h_c^2 - 1)}. \quad (63)$$

Note that we need $h_c^4 \geq h_d^2 + h_c^2$ since $\theta \in [0, 1]$. We can then rewrite $Y_1$ as

$$
Y_1 = \frac{h_d}{h_c^2} \sqrt{h_c^4 - h_c^2 - h_d^2} U_{1a} + \frac{h_d^2}{h_c^2} U_{1b} + \frac{\sqrt{h_c^4 - h_c^2 - h_d^2}}{h_c} U_{2a}
$$
$$
+ \frac{h_d}{h_c}(V_1 + U_{2b}) + Z_1 + V_2. \tag{64}
$$

Since the channel is symmetric, $Y_2$ is similarly obtained as

$$
Y_2 = \frac{h_d}{h_c^2} \sqrt{h_c^4 - h_c^2 - h_d^2} U_{2a} + \frac{h_d^2}{h_c^2} U_{2b} + \frac{\sqrt{h_c^4 - h_c^2 - h_d^2}}{h_c} U_{1a}
$$
$$
+ \frac{h_d}{h_c}(V_2 + U_{1b}) + Z_2 + V_1. \tag{65}
$$

At the same time, the sources also utilize the OBRC to send the messages $W_{icb}$ using a nested lattice code following the construction in section IV-A.2. Specifically, the sources map $W_{icb}$ to lattice point $t_{icb}^m \in \Lambda_c \cap \mathcal{V}(\Lambda)$, and transmit

$$
X_{iR}^m = (t_{icb}^m + D_i^m) \mod \Lambda, \tag{66}
$$

where $D_i^m \sim \text{Unif}(\mathcal{V}(\Lambda))$ is the dither. The relay decodes the modulo sum of these two messages $t_{1cb}^m \oplus t_{2cb}^m$, and forwards it to the destinations.

The decoder at destination 1 decodes the signal transmitted through the IC in the following order: First the signal $U_{1a}^m$ is decoded, followed by $U_{1b}^m$ and $U_{2a}^m$. The decoder also decodes the signal transmitted through the OBRC to obtain $t_{1cb}^m \oplus t_{2cb}^m$. Since decoder 1 knows $W_{1cb}$ from decoding $U_{1b}$, it can recover $W_{2cb}$ from $t_{1cb}^m \oplus t_{2cb}^m$. The interference signal $U_{2b}$ then can be subtracted from $Y_1$, and $V_1$ can be decoded. To guarantee each step has vanishing error probability, we need the following inequalities to be satisfied.

$$
R_{1ca} \leq \frac{1}{2} \log \left( 1 + \frac{h_d^2 (h_c^4 - h_c^2 - h_d^2)}{h_c^6 + h_c^4 + h_d^4 + h_d^2 h_c^2} \right) \tag{67}
$$

$$
R_{1cb} \leq \frac{1}{2} \log \left( 1 + \frac{h_d^4}{h_c^6 + h_c^4 + h_d^2 h_c^2} \right) \tag{68}
$$

$$
R_{1cb} = R_{2cb} \leq \left( \frac{1}{4} \log \left( \frac{1}{2} + h_r^2 \right) \right)^+ \tag{69}
$$

$$
R_{2ca} \leq \frac{1}{2} \log \left( 1 + \frac{h_c^4 - h_c^2 - h_d^2}{2h_d^2 + 2h_c^2} \right) \tag{70}
$$

$$
R_{1p} \leq \frac{1}{2} \log \left( 1 + \frac{h_d^2}{2h_c^2} \right). \tag{71}
$$

The decoder at destination 2 is identical to above and the rate constraints at destination 2 can be obtained by switching the indices 1 and 2. It is easy to verify that (67) is larger than (70) since $h_d^2(h_d^2 + h_c^2) > h_c^4(h_c^2 + 1)$. When the condition (57) holds, the sum rate (58) is achievable.

It can be verified that the gap between this rate and the outerbound $C_{sum,3}$ in (24) is at most $\log 2\sqrt{6} = 2.2925$ bits, see Appendix C.

*Relation to the deterministic model*: The condition (57) corresponds to $n_r < 4n_d - 6n_c$ for the deterministic model, which means that all resources in OBRC are used to help the destinations decode part of the common interference messages and no resource can be utilized to send new information. When (57) does not hold, the achievable scheme can be further improved by sending new messages $W_{iR}$ in addition to $W_{icb}$ through the OBRC following the steps (48)-(39).

Using this approach, we can show that the sum rate (59) is achievable where we choose the parameter $\gamma$ such that the rate constraints for $W_{icb}$ are the same at the IC and the OBRC, which requires the condition (60).

It can be verified that the gap between this rate and the outerbound $C_{sum,4}$ in (25) is at most 2.25 bits, see Appendix C. ∎

*Proposition 8:* For **very weak interference:** $h_c^4 < h_d^2 + h_c^2$, when the following condition holds

$$
\left( \frac{1}{4} \log \left( \frac{1}{2} + h_r^2 \right) \right)^+ \leq \frac{1}{2} \log \left( 1 + \frac{h_d^2(h_c^2 - 1)}{h_c^4 + h_c^2 + h_d^2} \right), \tag{72}
$$

the following sum rate is achievable

$$
R_{sum} = \log \left( 1 + \frac{h_d^2}{2h_c^2} \right) + \left( \frac{1}{2} \log \left( \frac{1}{2} + h_r^2 \right) \right)^+. \tag{73}
$$

Otherwise, the following sum rate is achievable

$$
R_{sum} = \log \left( 1 + \frac{h_d^2}{2h_c^2} \right) + \log \left( 1 + \frac{h_d^2(h_c^2 - 1)}{h_c^4 + h_c^2 + h_d^2} \right)
$$
$$
+ \frac{1}{4} \log \left( \frac{1 + h_r^2}{1 + h_r^2 \gamma^2} \right), \tag{74}
$$

where the parameter $\gamma$ is chosen such that the following condition holds

$$
\frac{1}{2} \log \left( 1 + \frac{h_d^2(h_c^2 - 1)}{h_c^4 + h_c^2 + h_d^2} \right) = \frac{1}{4} \log \left( \frac{1}{2} + h_r^2 \gamma^2 \right). \tag{75}
$$

Moreover, the above sum rates (73) and (74) have constant gap with the outerbounds.

*Proof:* Note that the rate splitting strategy for the case of weak interference does not work for this range of channel parameters, since it requires $h_c^4 \geq h_d^2 + h_c^2$. From the insights obtained from the deterministic model, i.e., Figure 11 and the strategies described in Section III-F, we observe that it is sufficient to split the messages into two parts, i.e., common and private parts, where the private part is aligned at the noise level at the interference link.

Based on this insight, the sources split the message $W_i$ into common part $W_{ic}$ and private part $W_{ip}$. The sources further encode $W_{ic}$ and $W_{ip}$ into $U_i^m$ and $V_i^m$ respectively, where $U_i, V_i \sim \mathcal{N}(0, 1)$. The signal transmitted into the IC is

$$
X_i = \beta U_i + \sqrt{1 - \beta^2} V_i. \tag{76}
$$

We choose the parameter $\beta$ such that $V_i$ arrives at noise level at the interference links, i.e., $\beta^2 = \frac{h_c^2 - 1}{h_c^2}$.

The signal received at destination 1 is

$$
Y_1 = \frac{h_d}{h_c} \sqrt{h_c^2 - 1} U_1 + \frac{h_d}{h_c} V_1
$$
$$
+ \sqrt{h_c^2 - 1} U_2 + Z_1 + V_2. \tag{77}
$$

For the OBRC, the sources map $W_{ic}$ into lattice points following the construction in section IV-A.2. The relay decodes the modulo sum of the lattice points based on the signal received from its incoming bands, and then transmits the modulo sum to the destinations.

Destination 1 first decodes $W_{1c}$ from the signals received from the IC, and then recovers $W_{2c}$ from the signals received from the OBRC. Therefore, the interference signal $U_2$ can be removed, and $V_1$ can be decoded. To guarantee vanishing error probability for each decoding step, we need the following rate constraints at destination 1:

$$R_{1c} \leq \frac{1}{2} \log \left(1 + \frac{h_d^2(h_c^2 - 1)}{h_c^4 + h_c^2 + h_d^2}\right) \tag{78}$$

$$R_{1c} = R_{2c} \leq \left(\frac{1}{4} \log \left(\frac{1}{2} + h_r^2\right)\right)^+ \tag{79}$$

$$R_{1p} \leq \frac{1}{2} \log \left(1 + \frac{h_d^2}{2h_c^2}\right). \tag{80}$$

Destination 2 uses the same decoder, and the rate constraints at destination 2 can be obtained by switching indices 1 and 2 in the above rate expressions. We can show that the rate (73) is achievable when the condition (72) holds.

It can be verified that the gap between this rate and the outerbound $C_{sum,3}$ in (24) is at most 2.2925 bits, see Appendix C.

*Relation with the deterministic model:* The condition (72) corresponds to the condition $n_r < 2n_c$ in the deterministic model, where all resources of the OBRC are used to decode the common interference message. When the condition (72) does not hold, or $n_r \geq 2n_c$, the achievable sum rate can be improved by transmitting new message $W_{iR}$ in addition to $W_{ic}$ through the OBRC following the steps (48)-(39).

It is then easy to verify that the sum rate (74) is also achievable, and the parameter $\gamma$ guarantees that the rate constraints for the message $W_{ic}$ are the same at the IC and the OBRC, i.e., condition (75) holds.

We can show that that the gap between this rate and $C_{sum,4}$ in (25) is at most 1.75 bits. ∎

### C. Constant gap result for symmetric capacity

Based on the derivations in Section IV-A and IV-B, we conclude that our achievable strategy achieves within 1.14625 bits of the symmetric capacity, $C_{sym}$, for $h_c \geq 1$ and $h_d \geq 1$, since $C_{sum} = 2C_{sym}$. When $h_c < 1$ or $h_d < 1$, we can apply the same strategies used in the cases when $h_c \geq 1$ and $h_d \geq 1$ by treating the signals coming from links with strength less than 1 as noise. For example, when interference is very strong or extremely strong but $h_d < 1$, there is no need to split the messages. We replace (42) with

$$Y_1 = h_c X_2 + Z_1 + Z_{X_1}, \tag{81}$$

where $Z_{X_1} = h_d X_1$. The destinations do not decode any message from the direct links. Instead, the non-intended messages are decoded from the interference links first, and then the OBRC is utilized to recover the source messages. When interference is weak or very weak but $h_c < 1$, it is sufficient to use separate encoding between the IC and the OBRC.

We replace (64) with

$$Y_1 = h_d X_1 + Z_1 + Z_{X_2}, \tag{82}$$

where $Z_{X_2} = h_c X_2$. The destinations first decode the intended source message treating interference as noise, and then decode the signals transmitted through the OBRC from both sources, as in the multiple-access channel. We can show that the same constant gap results hold for $h_c < 1$ or $h_d < 1$. This leads to the result in the title of this paper:

*Theorem 2:* The symmetric capacity $C_{sym}$ of the symmetric Gaussian IC-OBR is within 1.14625 bits of $\overline{C}_{sym}$ for fixed duplexing factor 0.5, and is within 1.27125 bits of $\overline{C}_{sym}$ for arbitrary duplexing factors, i.e., for $\alpha = 0.5$,

$$\overline{C}_{sym} - 1.14625 \leq C_{sym} \leq \overline{C}_{sym}, \tag{83}$$

for arbitrary $\alpha$,

$$\overline{C}_{sym} - 1.27125 \leq C_{sym} \leq \overline{C}_{sym}, \tag{84}$$

where

$$\overline{C}_{sym} = \frac{1}{2} \min\{C_{sum,1}, C_{sum,2}\} \tag{85}$$

for $h_c^2 \geq h_d^2$, and

$$\overline{C}_{sym} = \frac{1}{2} \min\{C_{sum,3}, C_{sum,4}\} \tag{86}$$

for $h_c^2 < h_d^2$.

### D. A case when interference is useful

From *Theorem 2*, an important observation is that strong interference can potentially improve capacity when an OBR is present in the system. To justify this observation, we first assume that there is no interference in the model, i.e., $h_c = 0$. The upperbound for the symmetric capacity of IC-OBR is then

$$\frac{1}{2} \log \left(1 + h_d^2\right) + \frac{1}{8} \log \left(1 + h_r^2\right) \tag{87}$$

for fixed $\alpha = 0.5$. When interference link is extremely strong, we are able to achieve within 0.65 bits of the rate

$$\frac{1}{2} \log(1 + h_d^2) + \frac{1}{4} \log(1 + h_r^2), \tag{88}$$

which can be larger than the upperbound for the capacity of IC-OBR without interference when

$$\log(1 + h_r^2) > 5.2. \tag{89}$$

When there is no OBR, the benefit of strong interference is that it does not reduce the rate [3], i.e., interference at most has a *neutral* effect. With OBR, however, we can see that the strong interference can further improve the rates, and thus it has a *positive* effect on the capacity.

### E. Discussion

*Theorem 2* and the achievable strategies developed leading to it provide us with insights as to how to handle the interference with an OBR. For extremely strong and very strong interference, the interference links support much larger rates than the direct links. For the IC without OBR, the excessive rates of the interference links can only help with interference cancelation to achieve the maximum rates supported by the direct links. When the OBR is added to the system, the interference links can be used to convey side information to the destinations. This side information can facilitate the transmission through the OBRC. In particular, we observe that when the interference is extremely strong, the channel acts as if there are *two disjoint OBRC assisting each source-destination pair*. This can be seen from the first term in (85), which is

$$\frac{1}{2}\log(1 + h_d^2) + \frac{1}{4}\log(1 + h_r^2). \qquad (90)$$

Under the condition (36), both users can achieve within 0.65 bits of this rate. The term $\frac{1}{4}\log(1 + h_r^2)$ acts as if there are two independent OBRCs, one for each source-destination pair.

When interference link is weaker than the direct link, the HK strategy splits the message into common and private, where the common message causes interference at the non-intended receivers. Recall that we term the common message from the intended source as the common information message, while the common message from the non-intended source as the common interference message, from the receiver's perspective. Without the OBRC, the decoder needs to decode both the common interference and information messages to reduce the effect of the interference. This approach is shown to achieve within 1 bit of the capacity for the IC without OBR [7]. For IC-OBR, applying this approach for the IC with separate encoding for the OBRC only works well in moderate interference. For weak and very weak interference, it has unbounded gap with the outerbounds.

The OBRC, in effect, provides a new vehicle to handle the interference. Note that in our strategy, the common information messages are always decoded from the signals obtained from the IC under weak and very weak interference. The common interference messages, on the other hand, need to be treated in a smarter fashion in order to improve the achievable rates. For weak interference, it is beneficial to decode part of the common interference message from the signals obtained from the IC, while using the OBRC to recover the rest of the common interference message. However, when the interference is very weak, the decoder should not decode any common interference message from the signals obtained from the IC. To achieve higher rates, it should recover all the common interference message from the signals obtained from the OBRC.

To see why these approaches work well for the IC-OBR, we first examine the case when the interference is very weak. For this case, decoding all parts of the common interference message at the non-intended receiver imposes a severe constraint on the rate of the message. As we recall from *Proposition 6*, when the destinations decode both the common interference and information messages from the IC, and use the OBRC to transmit new information, the achievable sum rate is

$$\begin{aligned}
R_{sum} = {} & \log\left(1 + \frac{h_d^2}{2h_c^2}\right) + \frac{1}{4}\log\left(1 + h_r^2\right) \\
& + \min\left\{\frac{1}{2}\log\left(1 + \frac{(h_d^2 + h_c^2)(h_c^2 - 1)}{2h_c^2 + h_d^2}\right), \right. \\
& \left. \qquad\qquad \log\left(1 + \frac{h_c^2(h_c^2 - 1)}{2h_c^2 + h_d^2}\right)\right\}. \quad (91)
\end{aligned}$$

We can see that the rate expression

$$\log\left(1 + \frac{h_c^2(h_c^2 - 1)}{2h_c^2 + h_d^2}\right) \qquad (92)$$

is due to decoding the common interference message. Clearly this rate is limited when $h_c^4 < h_d^2 + h_c^2$. Under this condition, it is also easy to verify that the gap between this rate and outerbound $C_{sum,3}, C_{sum,4}$ is unbounded. When we use the above approach to decode the common interference messages from the OBRC, the rate constraint (92) can be relaxed, and the resulting achievable rate has a constant gap with the outerbounds.

Nevertheless, for $h_d^2(h_d^2 + h_c^2) > h_c^4(h_c^2 + 1)$, and $h_c^4 \geq (h_d^2 + h_c^2)$, decoding all the common interference message from the OBRC cannot achieve within constant gap of the outerbounds. In this case, it is beneficial to further split the common messages into two parts, and decode one part of the common interference message using the IC and the other part using the OBRC, since the interference now is stronger than the previous case when $h_c^4 < h_d^2 + h_c^2$.

The reason for further splitting the common messages can be better illustrated using the deterministic model. From Figure 9, we can see that if we do not split the common messages, the sources need to transmit the signal bits from $A_1, A_2$ and $B_1, B_2$ using the OBRC. Destination 1 first decodes the signals from $A_1, A_2$ using the IC, and then use the OBRC to obtain the signal bits from $B_1, B_2$ to remove the common interference messages. However, the signal bits from $B_1$ can be decoded directly from the IC. Similar arguments hold for destination 2. Clearly this is suboptimal since the resources of the OBRC are not fully utilized, since the resources of the OBRC, which the sources used to transmit signal bits from $A_1$ and $B_1$, can be used to transmit new information bits. Therefore, further splitting the messages is needed. The constant gap result shows the advantage of this approach.

## V. CONCLUSION

In this paper, we have studied the deterministic IC-OBR and established its sum capacity results for all possible channel parameters by deriving new outerbounds and constructing achievable strategies. We have also studied the Gaussian IC-OBR and established a constant gap result for the symmetric capacity.

We have classified the interference links as extremely strong, very strong, strong, moderate, weak, and very weak, according to the relative strength between the interference links, direct links, and links in the OBRC. By deriving outerbounds and constructing achievable strategies, we have shown that separate encoding is good for strong and moderate interference. We have also shown that for very strong and extremely strong interference, the interference links can convey some side information to the non-intended receivers, which can be used by the OBRC to transmit additional messages. For weak and very weak interference, we have shown that the OBRC plays an important role in decoding the common messages, which improves the achievable rates. We have shown that the achievable strategies proposed in this paper achieve the symmetric capacity to within 1.14625 bits for fixed duplexing factor 0.5, or 1.27125 bits for arbitrary duplexing factors. An important observation from the constant gap result is that strong interference can be useful to improve the achievable rates with the presence of an OBR.

The results in this paper provide us with insights as to how to utilize and manage interference using relay nodes in interference limited wireless networks.

# APPENDIX A
# OUTERBOUNDS FOR THE DETERMINISTIC IC-OBR

## A. Optimal Duplexing Factor for the Outerbounds

We first show that the optimal duplexing factor $\alpha^*$ cannot be less than 0.5. We prove by contradiction and suppose $\alpha^* < 0.5$. Since we are considering the deterministic model in the symmetric setting, the signal bits received at the relay can be forwarded to the destinations in a lossless manner. Therefore any encoding/decoding function performed at the relay can be deferred to the destinations. It is then equivalent to consider the following scenario: the relay listens to the channel for $\alpha^* m$ channel uses, and it uses another $\alpha^* m$ channel uses to transmit the original signal bits it received to the destinations. Any transmission using $\alpha^* < 0.5$ thus can be improved by using $\alpha^* = 0.5$. We conclude that $0.5 \leq \alpha^* \leq 1$.

Next, we can bound the sum rate as

$$m(R_1 + R_2) \tag{93}$$
$$\leq I(W_1; \mathbf{y}_{11}^m, \mathbf{y}_{12,\alpha m+1}^m) + I(W_2; \mathbf{y}_{21}^m, \mathbf{y}_{22,\alpha m+1}^m) \tag{94}$$
$$\leq I(W_1; \mathbf{y}_{11}^m, \mathbf{y}_{12,\alpha m+1}^m, \mathbf{y}_{12,0.5m+1}^{\alpha m}) $$
$$\quad + I(W_2; \mathbf{y}_{21}^m, \mathbf{y}_{22,\alpha m+1}^m, \mathbf{y}_{22,0.5m+1}^{\alpha m}) \tag{95}$$
$$= I(W_1; \mathbf{y}_{11}^m, \mathbf{y}_{12,0.5m+1}^m) + I(W_2; \mathbf{y}_{21}^m, \mathbf{y}_{22,0.5m+1}^m). \tag{96}$$

We can bound the individual rate in the same fashion. Therefore, we conclude that for the outerbound, 0.5 is the optimal duplexing factor.

## B. Sum Rate Outerbounds

The bounds (18) and (19) follow from the cut set bound. We now derive the rest of the sum rate bounds (20)-(21).

When $n_c \geq n_d$, we have

$$2m(R_1 + R_2)$$
$$= H(W_1) + H(W_2) \tag{97}$$
$$= I(W_1; \mathbf{y}_{11}^{2m}, \mathbf{y}_{12}^m) + H(W_1|\mathbf{y}_{11}^{2m}, \mathbf{y}_{12}^m) + I(W_2; \mathbf{y}_{21}^{2m}, \mathbf{y}_{22}^m)$$
$$\quad + H(W_2|\mathbf{y}_{21}^{2m}, \mathbf{y}_{22}^m) \tag{98}$$
$$\leq I(\mathbf{x}_{11}^{2m}, \mathbf{x}_{12}^{2m}; \mathbf{y}_{11}^{2m}, \mathbf{y}_{12}^m) + 2m\epsilon_1 + 2m\epsilon_2 $$
$$\quad + I(\mathbf{x}_{21}^{2m}, \mathbf{x}_{22}^m, \mathbf{x}_r^m; \mathbf{y}_{21}^{2m}, \mathbf{y}_{22}^m | \mathbf{x}_{11}^{2m}, \mathbf{x}_{12}^m) \tag{99}$$
$$= H(\mathbf{y}_{11}^{2m}, \mathbf{y}_{12}^m) - H(\mathbf{y}_{11}^{2m}, \mathbf{y}_{12}^m | \mathbf{x}_{11}^{2m}, \mathbf{x}_{12}^m)$$
$$\quad + H(\mathbf{y}_{21}^{2m}, \mathbf{y}_{22}^m | \mathbf{x}_{11}^{2m}, \mathbf{x}_{12}^m) \tag{100}$$
$$= H(\mathbf{y}_{11}^{2m}, \mathbf{y}_{12}^m) - H(\mathbf{S}^{q-n_c} \mathbf{x}_{21}, \mathbf{y}_{12}^m | \mathbf{x}_{11}^{2m}, \mathbf{x}_{12}^m)$$
$$\quad + H(\mathbf{S}^{q-n_d} \mathbf{x}_{21}, \mathbf{y}_{22}^m | \mathbf{x}_{11}^{2m}, \mathbf{x}_{12}^m) \tag{101}$$
$$\leq H(\mathbf{y}_{11}^{2m}, \mathbf{y}_{12}^m) \tag{102}$$
$$\leq 2m \cdot n_c + m \cdot n_r \tag{103}$$

where (102) is due to the symmetry of the channel model and the fact that $n_c \geq n_d$. Note that we use superscript $2m$ for signal received from the IC and $m$ for signal received from the OBRC, since we are using duplexing factor 0.5. We can then write the sum rate outerbound as

$$R_1 + R_2 \leq n_c + \frac{1}{2} n_r. \tag{104}$$

When $n_d > n_c$,

$$2m(R_1 + R_2) \tag{105}$$
$$= H(W_1) + H(W_2) \tag{106}$$
$$\leq I(\mathbf{x}_{11}^{2m}, \mathbf{x}_{12}^m; \mathbf{y}_{11}^{2m}, \mathbf{y}_{12}^m, \mathbf{v}_{11}^{2m}) + I(\mathbf{x}_{21}^{2m}, \mathbf{x}_{22}^m; \mathbf{y}_{21}^{2m}, \mathbf{y}_{22}^m, \mathbf{v}_{21}^{2m}) \tag{107}$$
$$= H(\mathbf{v}_{11}^{2m}) + H(\mathbf{y}_{11}^{2m}, \mathbf{y}_{12}^m | \mathbf{v}_{11}^{2m}) - H(\mathbf{v}_{21}^{2m}, \mathbf{y}_{12}^m | \mathbf{x}_{11}^{2m}, \mathbf{x}_{12}^m)$$
$$\quad + H(\mathbf{v}_{21}^{2m}) + H(\mathbf{y}_{21}^{2m}, \mathbf{y}_{22}^m | \mathbf{v}_{21}^{2m}) - H(\mathbf{v}_{11}^{2m}, \mathbf{y}_{22}^m | \mathbf{x}_{21}^{2m}, \mathbf{x}_{22}^m) \tag{108}$$
$$\leq H(\mathbf{y}_{11}^{2m}, \mathbf{y}_{12}^m | \mathbf{v}_{11}^{2m}) + H(\mathbf{y}_{21}^{2m}, \mathbf{y}_{22}^m | \mathbf{v}_{21}^{2m}) \tag{109}$$
$$\leq 2m \cdot n_r + 4m \max\{n_d - n_c, n_c\} \tag{110}$$

where $\mathbf{v}_{11} = \mathbf{S}^{q-n_c} \mathbf{x}_{11}$, $\mathbf{v}_{21} = \mathbf{S}^{q-n_c} \mathbf{x}_{21}$ are the genie information we give to the decoders. The sum rate outerbound can be written as

$$R_1 + R_2 \leq n_r + 2 \max\{n_d - n_c, n_c\}. \tag{111}$$

Next, we can use another method to bound the sum rate, which is similar to the one in [23].

$$2m(R_1 + R_2)$$
$$= H(W_1) + H(W_2) \tag{112}$$
$$= H(W_1, W_2) \tag{113}$$
$$= I(W_1, W_2; \mathbf{y}_{11}^{2m}, \mathbf{y}_{12}^m, \mathbf{u}_{21}^{2m}) + H(W_1, W_2 | \mathbf{y}_{11}^{2m}, \mathbf{y}_{12}^m, \mathbf{u}_{21}^{2m}) \tag{114}$$
$$\leq I(\mathbf{x}_{11}^{2m}, \mathbf{x}_{12}^m, \mathbf{x}_{21}^{2m}, \mathbf{x}_{22}^m, \mathbf{x}_r^m; \mathbf{y}_{11}^{2m}, \mathbf{y}_{12}^m, \mathbf{u}_{21}^{2m})$$
$$\quad + H(W_1 | \mathbf{y}_{11}^{2m}, \mathbf{y}_{12}^m) + H(W_2 | \mathbf{y}_{11}^{2m}, \mathbf{y}_{12}^m, \mathbf{u}_{21}^{2m}, W_1) \tag{115}$$
$$\leq I(\mathbf{x}_{11}^{2m}, \mathbf{x}_{12}^m, \mathbf{x}_{21}^{2m}, \mathbf{x}_{22}^m, \mathbf{x}_r^m; \mathbf{y}_{11}^{2m}, \mathbf{y}_{12}^m, \mathbf{u}_{21}^{2m}) + 2m\epsilon_1$$
$$\quad + H(W_2 | \mathbf{y}_{11}^{2m}, \mathbf{y}_{12}^m, \mathbf{u}_{21}^{2m}, W_1, \mathbf{x}_{11}^{2m}, \mathbf{x}_{12}^m) \tag{116}$$
$$\leq I(\mathbf{x}_{11}^{2m}, \mathbf{x}_{12}^m, \mathbf{x}_{21}^{2m}, \mathbf{x}_{22}^m, \mathbf{x}_r^m; \mathbf{y}_{11}^{2m}, \mathbf{y}_{12}^m, \mathbf{u}_{21}^{2m}) + 2m\epsilon_1$$
$$\quad + H(W_2 | \mathbf{y}_{21}^{2m}, \mathbf{y}_{11}^{2m}, \mathbf{y}_{12}^m, \mathbf{u}_{21}^{2m}, W_1, \mathbf{x}_{11}^{2m}, \mathbf{x}_{12}^m) \tag{117}$$
$$\leq I(\mathbf{x}_{11}^{2m}, \mathbf{x}_{12}^m, \mathbf{x}_{21}^{2m}, \mathbf{x}_{22}^m, \mathbf{x}_r^m; \mathbf{y}_{11}^{2m}, \mathbf{y}_{12}^m, \mathbf{u}_{21}^{2m}) + 2m\epsilon_1$$
$$\quad + H(W_2 | \mathbf{y}_{21}^{2m}, \mathbf{y}_{22}^m) \tag{118}$$
$$\leq H(\mathbf{y}_{11}^{2m}, \mathbf{y}_{12}^m, \mathbf{u}_{21}^{2m}) + 2m\epsilon_1 + 2m\epsilon_2 \tag{119}$$
$$\leq 2m \cdot n_d + m \cdot n_r + 2m \cdot (n_d - n_c) \tag{120}$$

where $\mathbf{u}_{21} = [\mathbf{S}^{q-n_d} \mathbf{x}_{21}]^{\uparrow n_c}$ is the genie information we give to the decoder 1, and $\mathbf{x}^{\uparrow n_c}$ denotes the operation of removing the first $n_c$ elements of the vector $\mathbf{x}$. The step (116) is because $\mathbf{x}_{11}^{2m}$ and $\mathbf{x}_{12}^m$ are functions of $W_1$, and step (117) is because given $\mathbf{y}_{11}^{2m}, \mathbf{u}_{21}^{2m}, \mathbf{x}_{11}^{2m}$, we can recover $\mathbf{y}_{21}^{2m}$, and $\mathbf{y}_{22}^m = \mathbf{y}_{12}^m$. The sum rate upperbound is

$$R_1 + R_2 \leq 2n_d - n_c + \frac{1}{2} n_r. \tag{121}$$

Combining the two terms yields the result:

$$R_1 + R_2 \leq \min\{n_r + 2 \max\{n_d - n_c, n_c\}, 2n_d - n_c + \frac{1}{2} n_r\}. \tag{122}$$

# APPENDIX B
# OUTERBOUNDS FOR THE GAUSSIAN IC-OBR

The bound $C_{sum,1}$ is the sum of the individual rates from the cut set bounds with cuts at the relay and destinations. The bound $C_{sum,2}$ can be obtained using the strong interference condition and the symmetry of the channel with an argument similar to the one used in the strong interference channel in [5], along with the fact that we can obtain two outerbounds: one using output at destination for OBRC, and the other one using output at the relay for OBRC.

The bound $C_{sum,3}$ is obtained using genie argument. Specifically, we have

$$m(R_1 + R_2) \tag{123}$$
$$\leq I(W_1; Y_1^m, Y_{1R,\alpha m+1}^m) + I(W_2; Y_2^m, Y_{2R,\alpha m+1}^m) \tag{124}$$
$$= I(W_1; Y_1^m) + I(W_2; Y_2^m) + I(W_1; Y_{1R,\alpha m+1}^m | Y_1^m)$$
$$\quad + I(W_2; Y_{2R,\alpha m+1}^m | Y_2^m) \tag{125}$$
$$\leq I(W_1; Y_1^m, h_c X_1^m + Z_1^m) + I(W_2; Y_2^m, h_c X_2^m + Z_2^m)$$
$$\quad + I(W_1 X_{R,\alpha m+1}^m; Y_{1R,\alpha m+1}^m | Y_1^m)$$
$$\quad + I(W_2 X_{R,\alpha m+1}^m; Y_{2R,\alpha m+1}^m | Y_2^m) \tag{126}$$
$$\leq h(Y_1^m | h_c X_1^m + Z_1^m) - h(Z_1^m) + h(Y_2^m | h_c X_2^m + Z_2^m)$$
$$\quad - h(Z_2^m) + h(Y_{1R,\alpha m+1}^m)$$
$$\quad + h(Y_{2R,\alpha m+1}^m) - h(Z_{2R,\alpha m+1}^m) \tag{127}$$
$$\leq \log\left(1 + h_c^2 + \frac{h_d^2}{1 + h_c^2}\right) + (1 - \alpha)\log\left(1 + h_r^2\right). \tag{128}$$

The other part of $C_{sum,3}$ can be obtained by using $Y_R^{\alpha m}$ instead of $Y_{1R,\alpha m+1}^m$, i.e.,

$$m(R_1 + R_2) \tag{129}$$
$$\leq I(W_1; Y_1^m, Y_{1R,\alpha m+1}^m) + I(W_2; Y_2^m, Y_{2R,\alpha m+1}^m) \tag{130}$$
$$\leq I(W_1; Y_1^m, Y_R^{\alpha m}) + I(W_2; Y_2^m, Y_R^{\alpha m}), \tag{131}$$

since $X_{R,\alpha m+1}^m = f(Y_R^{\alpha m})$ and

$$Y_{iR,\alpha m+1}^m = h_r X_{R,\alpha m+1}^m + Z_{\alpha m+1}^m. \tag{132}$$

For the bound $C_{sum,4}$, we have

$$m(R_1 + R_2) \tag{133}$$
$$\leq I(W_1; Y_1^m, Y_{1R,\alpha m+1}^m | W_2) + I(W_2; Y_2^m, Y_{2R,\alpha m+1}^m) \tag{134}$$
$$= h(Y_1^m, Y_{1R,\alpha m+1}^m | W_2) - h(Y_1^m, Y_{1R,\alpha m+1}^m | W_1 W_2)$$
$$\quad + h(Y_2^m, Y_{2R,\alpha m+1}^m) - h(Y_2^m, Y_{2R,\alpha m+1}^m | W_2) \tag{135}$$
$$\leq h(h_d X_1^m + Z_1^m, Y_{1R,\alpha m+1}^m | W_2) - h(Z_1^m, Z_{1R,\alpha m+1}^m)$$
$$\quad + h(Y_2^m, Y_{2R,\alpha m+1}^m) - h(h_c X_1^m + Z_2^m, Y_{2R,\alpha m+1}^m | W_2) \tag{136}$$
$$= h(h_d X_1^m + Z_1^m) - h(h_c X_1^m + Z_2^m)$$
$$\quad + h(Y_{1R,\alpha m+1}^m | W_2, h_d X_1^m + Z_1^m)$$
$$\quad - h(Y_{2R,\alpha m+1}^m | h_c X_1^m + Z_2^m, W_2)$$
$$\quad + h(Y_2^m, Y_{2R,\alpha m+1}^m) - h(Z_1^m, Z_{1R,\alpha m+1}^m) \tag{137}$$
$$\leq h(h_d X_1^m + Z_1^m) - h(h_c X_1^m + Z_2^m)$$
$$\quad + h(Y_2^m, Y_{2R,\alpha m+1}^m) - h(Z_1^m, Z_{1R,\alpha m+1}^m) \tag{138}$$
$$\leq \frac{1}{2}\log\left(1 + h_d^2\right) + \frac{1}{2}\log\left(1 + \frac{h_d^2}{1 + h_c^2}\right)$$
$$\quad + \frac{1 - \alpha}{2}\log\left(1 + h_r^2\right) \tag{139}$$

where (138) is due to the symmetry of the channel and the fact that $h_c < h_d$. Similarly, the other term in $C_{sum,4}$ can be obtained by using $Y_R^{\alpha m}$ instead of $Y_{iR,\alpha m+1}^m$.

## Appendix C
## Constant Gap Between Achievable Rates and Outerbounds

### A. Very Strong Interference

The sum rate (38) can be bounded as follows

$$\log\left(1 + \frac{h_d^2 - 1}{2}\right) + \log\left(1 + \frac{h_c^2}{h_d^4 + h_d^2}\right)$$
$$\quad\quad\quad\quad + \frac{1}{4}\log\left(\frac{1 + h_r^2}{1 + h_r^2\gamma^2}\right) \tag{140}$$
$$= \frac{1}{2}\log\left(1 + 2h_d^2 + h_d^4\right)\left(\frac{h_c^2 + h_d^4 + h_d^2}{h_d^4 + h_d^2}\right) + \frac{1}{4}\log\left(1 + h_r^2\right)$$
$$\quad + \frac{1}{4}\log\left(\frac{1}{2} + h_r^2\gamma^2\right) - \frac{1}{4}\log\left(1 + h_r^2\gamma^2\right) - 1 \tag{141}$$
$$\geq -1.25 + \frac{1}{2}\log(1 + h_d^2 + h_c^2)\} + \frac{1}{4}\log(1 + h_r^2). \tag{142}$$

The gap between this rate and the bound $C_{sum,2}$ is thus at most 1.25 bits.

### B. Weak Interference

We can write the expression (58) as

$$\log\left(1 + \frac{h_d^2}{2h_c^2}\right) + \log\left(1 + \frac{h_c^4 - h_c^2 - h_d^2}{2h_d^2 + 2h_c^2}\right)$$
$$\quad\quad\quad\quad + \left(\frac{1}{2}\log\left(\frac{1}{2} + h_r^2\right)\right)^+ \tag{143}$$
$$= \log\left(\frac{2h_c^2 + h_d^2}{2h_c^2}\right)\left(\frac{h_d^4 + h_c^2 + h_d^2}{h_d^2 + h_c^2}\right) - 1$$
$$\quad + \left(\frac{1}{2}\log\left(\frac{1}{2} + h_r^2\right)\right)^+ \tag{144}$$
$$\geq \log\left(1 + h_c^2 + \frac{h_d^2}{h_c^2}\right) - 2 + \left(\frac{1}{2}\log\left(\frac{1}{2} + h_r^2\right)\right)^+. \tag{145}$$

It is now easy to see that the gap between this rate and the outerbound $C_{sum,3}$ is at most $\log 2\sqrt{6} = 2.2925$ bits.

The rate (59) can be written as

$$\log\left(1 + \frac{h_d^2}{2h_c^2}\right) + \log\left(\frac{h_c^4 + h_c^2 + h_d^2}{2h_d^2 + 2h_c^2}\right)$$
$$\quad + \log\left(\frac{h_d^4 + h_c^6 + h_c^4 + h_d^2 h_c^2}{h_c^6 + h_c^4 + h_d^4 h_c^2}\right) + \frac{1}{4}\log\left(1 + h_r^2\right)$$
$$\quad - \frac{1}{4}\log\left(\frac{1}{2} + h_r^2\gamma^2\right) - \frac{1}{4}\log\left(1 + \frac{0.5}{0.5 + h_r^2\gamma^2}\right) \tag{146}$$
$$\geq \log\left(1 + \frac{h_d^2}{2h_c^2}\right) + \log\left(\frac{h_c^4 + h_c^2 + h_d^2}{h_d^2 + h_c^2}\right) - \frac{5}{4}\log 2$$
$$\quad + \frac{1}{2}\log\left(\frac{h_d^4 + h_c^6 + h_c^4 + h_d^2 h_c^2}{h_c^6 + h_c^4 + h_d^2 h_c^2}\right) + \frac{1}{4}\log\left(1 + h_r^2\right) \tag{147}$$
$$= \log\left(1 + \frac{h_d^2}{2h_c^2}\right) + \frac{1}{4}\log\left(1 + h_r^2\right) - \frac{5}{4}\log 2$$

$$+ \frac{1}{2} \log \left( \frac{(h_c^4 + h_c^2 + h_d^2)^2}{(h_d^2 + h_c^2)^2} \cdot \frac{h_d^4 + h_c^6 + h_c^4 + h_d^2 h_c^2}{h_c^6 + h_c^4 + h_d^2 h_c^2} \right)$$
$$(148)$$

$$\geq \log \left( 1 + \frac{h_d^2}{2h_c^2} \right) + \frac{1}{2} \log \left( \frac{h_c^4 + h_c^2 + h_d^2}{h_c^2} \right)$$
$$+ \frac{1}{4} \log \left( 1 + h_r^2 \right) - \frac{5}{4} \log 2 \qquad (149)$$

$$= \frac{1}{2} \log \left( 1 + \frac{h_d^2}{2h_c^2} \right) + \frac{1}{2} \log \left( \frac{(h_c^4 + h_c^2 + h_d^2)(2h_c^2 + h_d^2)}{2h_c^4} \right)$$
$$+ \frac{1}{4} \log \left( 1 + h_r^2 \right) - \frac{5}{4} \log 2 \qquad (150)$$

$$\geq \frac{1}{2} \log \left( 1 + \frac{h_d^2}{2h_c^2} \right) + \frac{1}{2} \log \left( 1 + h_c^2 + \frac{1}{2} h_d^2 \right)$$
$$+ \frac{1}{4} \log \left( 1 + h_r^2 \right) - \frac{5}{4} \log 2 \qquad (151)$$

where (149) is due to $h_c^6 = h_c^2 h_d^4 > h_c^2 h_d^2$.

It is now easy to verify that the gap between this rate and the outerbound $C_{sum,4}$ is at most 2.25 bits.

### C. Very Weak Interference

For the rate (73), we can show that the gap with the outerbound $C_{sum,3}$ is

$$\log \left( \frac{1 + h_c^2 + \frac{h_d^2}{1+h_c^2}}{1 + \frac{h_d^2}{2h_c^2}} \right) + \frac{1}{2} \log \left( 1 + h_r^2 \right)$$
$$- \left( \frac{1}{2} \log \left( \frac{1}{2} + h_r^2 \right) \right)^+ \qquad (152)$$

$$= \log \left( \frac{2h_c^4 + 2h_c^2 + 2h_d^2}{2h_c^2 + h_d^2} \right) + \frac{1}{2} \log \left( 1 + h_r^2 \right)$$
$$- \left( \frac{1}{2} \log \left( \frac{1}{2} + h_r^2 \right) \right)^+ \qquad (153)$$

$$\leq \log \left( \frac{4h_c^2 + 4h_d^2}{2h_c^2 + h_d^2} \right) + \frac{1}{2} \log \left( 1 + h_r^2 \right)$$
$$- \left( \frac{1}{2} \log \left( \frac{1}{2} + h_r^2 \right) \right)^+ \qquad (154)$$

$$\leq \log 2\sqrt{6} = 2.2925 \text{ bits} \qquad (155)$$

where (154) is due to $h_c^4 < h_d^2 + h_c^2$.

The rate (74) can be written as

$$\log \left( 1 + \frac{h_d^2}{2h_c^2} \right) + \log \left( 1 + \frac{h_d^2(h_c^2 - 1)}{h_c^4 + h_c^2 + h_d^2} \right)$$
$$+ \frac{1}{4} \log \left( 1 + h_r^2 \right) - \frac{1}{4} \log \left( \frac{1}{2} + h_r^2 \gamma^2 \right)$$
$$- \frac{1}{4} \log \left( 1 + \frac{0.5}{0.5 + h_r^2 \gamma^2} \right) \qquad (156)$$

$$\geq \log \left( 1 + \frac{h_d^2}{2h_c^2} \right) + \frac{1}{2} \log \left( 1 + \frac{h_d^2(h_c^2 - 1)}{h_c^4 + h_c^2 + h_d^2} \right)$$
$$+ \frac{1}{4} \log \left( 1 + h_r^2 \right) - \frac{1}{4} \log 2 \qquad (157)$$

$$> \log \left( 1 + \frac{h_d^2}{2h_c^2} \right) + \frac{1}{2} \log \left( \frac{h_c^4 + h_c^2 + h_d^2 h_c^2}{2(h_c^2 + h_d^2)} \right)$$
$$+ \frac{1}{4} \log \left( 1 + h_r^2 \right) - \frac{1}{4} \log 2 \qquad (158)$$

$$> \frac{1}{2} \log \left( 1 + \frac{h_d^2}{2h_c^2} \right) + \frac{1}{2} \log \left( \frac{h_c^4 + h_c^2 + h_d^2 h_c^2}{h_c^2 + h_d^2} \cdot \frac{h_c^2 + h_d^2}{h_c^2} \right)$$
$$+ \frac{1}{4} \log \left( 1 + h_r^2 \right) - \frac{5}{4} \log 2 \qquad (159)$$

$$= \frac{1}{2} \log \left( 1 + \frac{h_d^2}{2h_c^2} \right) + \frac{1}{2} \log \left( 1 + h_c^2 + h_d^2 \right)$$
$$+ \frac{1}{4} \log \left( 1 + h_r^2 \right) - \frac{5}{4} \log 2 \qquad (160)$$

where (160) is due to $h_c^4 < h_d^2 + h_c^2$.

It is now easy to verify that the gap between this rate and the outerbound $C_{sum,4}$ is 1.75 bits.

**Ye Tian** (S'10) received the B.E. degree in information engineering from Beijing University of Posts and Telecommunications, Beijing, China, in 2007. He is currently pursuing the Ph.D. degree and is a graduate research assistant with the Department of Electrical Engineering at the Pennsylvania State University, University Park, PA since 2007. His research interests include network information theory, wireless communication, and resource allocation for wireless networks. He currently serves as a student co-chair of the IEEE Information Theory Society Student Committee.

**Aylin Yener** (S'91, M'00) received her two B.Sc. degrees, with honors, in Electrical and Electronics Engineering, and in Physics, from Boğaziçi University, Istanbul, Turkey, in 1991, and the M.S. and Ph.D. degrees in Electrical and Computer Engineering from Rutgers University, NJ, in 1994 and 2000, respectively. During her Ph.D. studies, she was with Wireless Information Network Laboratory (WINLAB) in the Department of Electrical and Computer Engineering at Rutgers University, NJ. From September 2000 to December 2001, she was with the Electrical Engineering and Computer Science Department, Lehigh University, PA, where she was a P.C. Rossin Assistant Professor. In January 2002, she joined the faculty of The Pennsylvania State University, University Park, where she was an Assistant Professor, then Associate Professor, and is currently Professor of Electrical Engineering. During the academic year 2008-2009, she was a Visiting Associate Professor with the Department of Electrical Engineering, Stanford University, Stanford CA. Her research interests are in information theory, communication theory and network science, with emphasis on fundamental limits of wireless ad hoc networks and information theoretic security.

Dr. Yener received the NSF CAREER award in 2003 and is a member of the team that received the DARPA Information Theory for Mobile Ad Hoc Networks (ITMANET) Young Investigator Team Award in 2006. In 2010, she received the Penn State Engineering Alumni Society (PSEAS) Outstanding Research Award, and the Best Paper award from the Communication Theory Symposium in IEEE International Conference on Communications (ICC). Her service to IEEE includes membership in the Technical Program Committees of various annual conferences since 2002. She chaired the Communications Track in the Asilomar Conference on Signals, Systems and Computers in 2005 and in 2008. She served as the Technical Program Co-Chair for the Communication Theory Symposium of the IEEE International Conference on Communications (ICC) 2009 and for the Wireless Communications Symposium of the IEEE International Conference on Communications (ICC) 2008. She currently serves on the editorial advisory board of the IEEE TRANSACTIONS ON WIRELESS COMMUNICATIONS and as an editor for the IEEE TRANSACTIONS ON COMMUNICATIONS. Her service to the IEEE Information Theory Society includes chairing the Student Committee 2007-2009. She is the cofounder of the Annual School of Information Theory in North America, and served as the general Co-Chair of the First Annual School of Information Theory that was held at Penn State University, University Park, in June 2008, the Second Annual School of Information Theory at Northwestern University, Evanston, IL, in August 2009, and the Third Annual School of Information Theory at University of Southern California, Los Angeles, in August 2010. She currently serves as the treasurer of the IEEE Information Theory Society.